\pdfoutput=1
\documentclass[sigplan,10pt,nonacm,review=false]{acmart}
\renewcommand\footnotetextcopyrightpermission[1]{}
\usepackage{amsmath}
\usepackage{tikz}
\usetikzlibrary{arrows.meta,positioning,calc,shapes.misc,decorations.pathreplacing}
\usepackage{pgfplots}
\pgfplotsset{compat=1.18}
\definecolor{wongorange}{HTML}{E69F00}
\definecolor{wongsky}{HTML}{56B4E9}
\definecolor{wonggreen}{HTML}{009E73}
\definecolor{wongviolet}{HTML}{CC79A7}
\definecolor{wongvermilion}{HTML}{D55E00}
\definecolor{wongyellow}{HTML}{F0E442}
\pgfplotsset{
  paper/.style={
    font=\footnotesize,
    label style={font=\footnotesize},
    tick label style={font=\footnotesize},
    legend style={font=\footnotesize, draw=none, fill=none, row sep=-1pt},
    legend cell align=left,
    tick style={black},
    axis line style={black},
    every axis plot/.append style={line width=1pt},
  },
}
\usepackage{booktabs}
\usepackage{xspace}
\newcommand{\dl}{DL\xspace}
\newcommand{\odl}{ODL\xspace}
\newcommand{\mpod}{M\textsubscript{pod}\xspace}
\newcommand{\clean}{\mathit{clean}}
\newcommand{\dirty}{\mathit{dirty}}
\makeatletter
\newif\if@podemailrow
\def\@mkauthors@iii{%
  \author@bx@wd=\textwidth\relax
  \advance\author@bx@wd by -\author@bx@sep\relax
  \ifnum\@ACM@authorsperrow>0\relax
    \divide\author@bx@wd by \@ACM@authorsperrow\relax
  \else
    \ifcase\num@authorgroups
    \relax
    \or
    \or \divide\author@bx@wd by \num@authorgroups\relax
    \or \divide\author@bx@wd by \num@authorgroups\relax
    \or \divide\author@bx@wd by 2\relax
    \else \divide\author@bx@wd by 3\relax
    \fi
  \fi
  \advance\author@bx@wd by -\author@bx@sep\relax
  \gdef\@currentauthors{}%
  \gdef\@currentaffiliation{}%
  \global\@podemailrowfalse
  \def\@author##1{\global\@podemailrowfalse\ifx\@currentauthors\@empty
    \gdef\@currentauthors{\par##1}%
  \else
    \g@addto@macro\@currentauthors{\qquad##1}%
  \fi
  \gdef\and{}}%
  \def\email##1##2{\ifx\@currentaffiliation\@empty
    \global\@podemailrowtrue
    \gdef\@currentaffiliation{\bgroup
      \mathchardef\UrlBreakPenalty=10000\nolinkurl{##2}\egroup}%
  \else
    \if@podemailrow
      \g@addto@macro\@currentaffiliation{\quad\bgroup
        \mathchardef\UrlBreakPenalty=10000\nolinkurl{##2}\egroup}%
    \else
      \global\@podemailrowtrue
      \g@addto@macro\@currentaffiliation{\par\bgroup
        \mathchardef\UrlBreakPenalty=10000\nolinkurl{##2}\egroup}%
    \fi
  \fi}%
  \def\affiliation##1##2{\global\@podemailrowfalse\ifx\@currentaffiliation\@empty
    \gdef\@currentaffiliation{%
      \setkeys{@ACM@affiliation@}{obeypunctuation=false}%
      \setkeys{@ACM@affiliation@}{##1}\@ACM@resetaffil
      ##2\@ACM@checkaffil}%
  \else
    \g@addto@macro\@currentaffiliation{\par
      \setkeys{@ACM@affiliation@}{obeypunctuation=false}%
      \setkeys{@ACM@affiliation@}{##1}\@ACM@resetaffil
      ##2\@ACM@checkaffil}%
  \fi
  \global\let\and\@typeset@author@bx}%
  \hsize=\textwidth
  \global\setbox\mktitle@bx=\vbox{\noindent
    \unvbox\mktitle@bx\par\medskip\leavevmode
    \lineskip=1pc\relax\centering\hspace*{-1em}%
    \addresses\let\and\@typeset@author@bx\and\par\bigskip}}
\makeatother

\IfFileExists{numbers.tex}{\newcommand{\AlBothRuns}{14\xspace}
\newcommand{\AlCapUnsafeM}{5\xspace}

\newcommand{\AlChainInstances}{two\xspace}

\newcommand{\AlChainRuns}{22\xspace}
\newcommand{\AlInstances}{14\xspace}
\newcommand{\AlKeptRuns}{46\xspace}

\newcommand{\AlLeakBothFirst}{32\xspace}
\newcommand{\AlLeakInstances}{13\xspace}
\newcommand{\AlLeakOrdFirst}{1,652\xspace}
\newcommand{\AlLeakSrcExhausted}{8\xspace}
\newcommand{\AlLeakSrcFirst}{144\xspace}
\newcommand{\AlLeaveInstances}{13\xspace}
\newcommand{\AlNoneLaterCapM}{20\xspace}
\newcommand{\AlNoneLaterRuns}{3\xspace}

\newcommand{\AlOrdLargestM}{16.1\xspace}

\newcommand{\AlPubUnsafeInstances}{13\xspace}
\newcommand{\AlRuns}{152\xspace}
\newcommand{\AlSentOnlyInstances}{4\xspace}
\newcommand{\AlSentOnlyUnsafe}{5,868\xspace}
\newcommand{\AlSentOnlyUnsafeInstances}{3\xspace}
\newcommand{\AlStatesM}{183\xspace}
\newcommand{\AlStopped}{20\xspace}
\newcommand{\AlTermBothFirst}{2,894\xspace}
\newcommand{\AlTermOrdFirst}{10,802\xspace}
\newcommand{\AlTermSrcFirst}{5,274\xspace}

\newcommand{\EsCxlZeroFlushDev}{370\xspace}
\newcommand{\EsCxlZeroFlushHost}{151\xspace}
\newcommand{\EsHiBreakEven}{2\xspace}

\newcommand{\EsHiOrdNsHiLoad}{227\xspace}

\newcommand{\EsHiOrdNsLoLoad}{153\xspace}

\newcommand{\EsHiSaveHalfPctLoad}{36\xspace}
\newcommand{\EsHiSaveHalfPctSteady}{27\xspace}

\newcommand{\EsHiSavePctLoad}{71\xspace}
\newcommand{\EsHiSavePctSteady}{55\xspace}

\newcommand{\EsHiSrcNsHiLoad}{789\xspace}

\newcommand{\EsHiSrcNsLoLoad}{532\xspace}

\newcommand{\EsLkExclUsHi}{11.12\xspace}
\newcommand{\EsLkExclUsLo}{0.13\xspace}
\newcommand{\EsLkExtraOneUsHi}{2.28\xspace}
\newcommand{\EsLkExtraOneUsLo}{0.02\xspace}

\newcommand{\EsLkExtraTwoUsHi}{1.75\xspace}
\newcommand{\EsLkExtraTwoUsLo}{0.02\xspace}
\newcommand{\EsLkMixes}{10\xspace}
\newcommand{\EsLkNoneUsHi}{9.36\xspace}
\newcommand{\EsLkNoneUsLo}{0.06\xspace}

\newcommand{\EsSwitchAdds}{220\xspace}

\newcommand{\EsWDirectHi}{394\xspace}
\newcommand{\EsWDirectLo}{266\xspace}
\newcommand{\EsWProtoHi}{540\xspace}
\newcommand{\EsWProtoLo}{494\xspace}
\newcommand{\EsWSwitchHi}{764\xspace}
\newcommand{\EsWSwitchLo}{486\xspace}
\newcommand{\EsXerxesCleanHi}{272\xspace}
\newcommand{\EsXerxesCleanLo}{241\xspace}

\newcommand{\EsXerxesDirtyLo}{418\xspace}

\newcommand{\HiBadlinkInstances}{5\xspace}
\newcommand{\HiBadlinkMax}{6,336\xspace}
\newcommand{\HiBadlinkMin}{163\xspace}
\newcommand{\HiBadlinkSame}{two\xspace}
\newcommand{\HiCapM}{2\xspace}

\newcommand{\HiCountOps}{1,000,000\xspace}
\newcommand{\HiCountSeeds}{5\xspace}
\newcommand{\HiDepChecks}{3,112,502\xspace}
\newcommand{\HiDepDl}{2,648\xspace}
\newcommand{\HiDepExhausted}{5\xspace}
\newcommand{\HiDepStopped}{4\xspace}
\newcommand{\HiFirstPairOdl}{163\xspace}

\newcommand{\HiInPlacePct}{71\xspace}
\newcommand{\HiInstances}{9\xspace}
\newcommand{\HiLinesShareLoadPct}{29\xspace}
\newcommand{\HiLinesShareSteadyPct}{45\xspace}
\newcommand{\HiNeverNone}{1,011\xspace}
\newcommand{\HiNoCrashMissing}{2\xspace}
\newcommand{\HiNoCrashRuns}{35\xspace}
\newcommand{\HiNoneOdlExt}{582\xspace}

\newcommand{\HiOrdLinesLoad}{0.58\xspace}
\newcommand{\HiOrdLinesSteady}{0.68\xspace}
\newcommand{\HiOrdLoad}{0.29\xspace}

\newcommand{\HiOrdSteadyRemove}{0.50\xspace}

\newcommand{\HiOrderPowerRuns}{9\xspace}
\newcommand{\HiOrderRuns}{12\xspace}
\newcommand{\HiPmChecks}{2,851,018\xspace}
\newcommand{\HiPmStopped}{4\xspace}
\newcommand{\HiResizes}{5\xspace}
\newcommand{\HiRuns}{80\xspace}
\newcommand{\HiSrcLoad}{2.00\xspace}
\newcommand{\HiSrcLoadResize}{2.75\xspace}
\newcommand{\HiSrcSteady}{1.50\xspace}
\newcommand{\HiStatesM}{46.3\xspace}
\newcommand{\HiThreeFirstPairOdl}{163\xspace}
\newcommand{\HiThreeNoneOdl}{1,146\xspace}
\newcommand{\HiThreeRuns}{20\xspace}
\newcommand{\LkLayoutBothRuns}{32\xspace}

\newcommand{\LkLayoutFwdRuns}{29\xspace}

\newcommand{\LkLayoutSepRuns}{10\xspace}
\newcommand{\LkLayoutSharedOneRuns}{6\xspace}
\newcommand{\LkLayoutSharedTwoRuns}{16\xspace}
}{}
\IfFileExists{numbers-locktable.tex}{\newcommand{\LkRuns}{400\xspace}
\newcommand{\LkCrashes}{317,084\xspace}
\newcommand{\LkReplicaRuns}{400\xspace}

\newcommand{\LkTpccHiUse}{20.4\xspace}

\newcommand{\LkHotRdAllOne}{52.2\xspace}

\newcommand{\LkHotRdAllUse}{78.1\xspace}

\newcommand{\LkTpccHiOneHits}{5\xspace}
\newcommand{\LkTpccHiOneCrashes}{6,036\xspace}
\newcommand{\LkTpccLoOneCrashes}{6,036\xspace}
\newcommand{\LkAbsentMinPct}{69\xspace}
\newcommand{\LkRandGapMaxPts}{1.4\xspace}
\newcommand{\LkTpccOneUp}{0.09\xspace}
\newcommand{\LkColdOneUp}{2.9\xspace}
\newcommand{\LkHotOneDn}{8.5\xspace}
\newcommand{\LkHotOneUp}{52.2\xspace}
\newcommand{\LkAllUseDn}{67.3\xspace}
\newcommand{\LkAllUseUp}{78.1\xspace}
\newcommand{\LkHotKeyPct}{7.1\xspace}
\newcommand{\LkColdKeyPct}{0.70\xspace}
\newcommand{\LkStoresPerRecordMin}{5.6\xspace}
\newcommand{\LkStoresPerRecordMax}{66.4\xspace}

\newcommand{\LkStoresMin}{0.75\xspace}
\newcommand{\LkStoresMax}{32.53\xspace}
\newcommand{\LkExclMin}{0.09\xspace}
\newcommand{\LkExclMax}{7.12\xspace}
\newcommand{\LkExclShareMinPct}{2\xspace}
\newcommand{\LkExclShareMaxPct}{37\xspace}
\newcommand{\LkPlacementGapMaxPts}{0.12\xspace}
\newcommand{\LkContrastMinLoPts}{0.47\xspace}
\newcommand{\LkExInstances}{10\xspace}
\newcommand{\LkExFwdTornInstances}{4\xspace}
\newcommand{\LkExStatesMin}{76\xspace}
\newcommand{\LkExStatesMax}{3,709\xspace}
\newcommand{\LkExStressRuns}{200,000\xspace}
\newcommand{\LkExStressNone}{526\xspace}
\newcommand{\LkExStressLock}{331\xspace}
\newcommand{\LkExStressData}{25\xspace}

\newcommand{\LkCachePairs}{400\xspace}
\newcommand{\LkFlRuns}{200\xspace}
\newcommand{\LkFlCrashes}{158,542\xspace}
\newcommand{\LkForcedNoneMin}{0.15\xspace}
\newcommand{\LkForcedNoneMax}{11.88\xspace}
\newcommand{\LkForcedExclMin}{0.15\xspace}
\newcommand{\LkForcedExclMax}{10.54\xspace}
\newcommand{\LkFlSavedUp}{0.38\xspace}
\newcommand{\LkFlSavedOneUp}{0.07\xspace}
\newcommand{\LkFlBeAllMin}{1.8\xspace}
\newcommand{\LkFlBeAllMax}{14.2\xspace}
\newcommand{\LkFlBeExclDn}{2.6\xspace}
\newcommand{\LkFlBatchStepsMin}{0.09\xspace}
\newcommand{\LkFlBatchStepsMax}{1.00\xspace}
\newcommand{\LkFlBatchMaxNamed}{12\xspace}

\newcommand{\LkFlExclKeptMinPct}{84\xspace}
\newcommand{\LkFlAllLeastAtTwo}{2\xspace}
\newcommand{\LkFlPairs}{20\xspace}
}{}

\title{CXL Coherence Writes Back What Surviving Hosts See, Leaving Flushes Only for Ordering}

\author{Harvey Dam}
\author{Ganesh Gopalakrishnan}
\affiliation{%
  \institution{Kahlert School of Computing, University of Utah}
  \city{Salt Lake City}
  \state{Utah}
  \country{USA}}
\email{harvey.dam@utah.edu}
\email{ganesh@cs.utah.edu}


\begin{document}

\begin{abstract}
Hosts that share memory over Compute Express Link (CXL) cache its lines, and a host that crashes loses the lines it has modified and not written back. Designs for such memory stop every host when one crashes, assume stored energy, or flush after each operation. We show that when a crashing host's clients die with it, the flushes a host crash requires are those that keep one line from reaching the memory device before a line it depends on, and not those that make a finished operation durable. The reason is a property that CXL requires of the fabric: a modified line is written back before another host accesses it. Whatever a surviving host has seen is then on the device, and our crash condition lets a crash drop operations that only the crashed host observed. We prove the claim for operations that take effect in a single store, and test it on small instances of models of three published designs, whose operations store several times. A lock table needs no flush to keep the locks of surviving hosts. A persistent hash index keeps every operation a survivor has observed, and in two workloads it writes fewer than half as many lines to the device as its published code does. An allocator's flush of each new reference is neither sufficient nor necessary for the survivors' safety: the allocator lacks ordering flushes, one of them when a reference changes hands.

\end{abstract}

\begin{teaserfigure}
\centering
\begin{tikzpicture}[>=Latex, font=\footnotesize,
    stepb/.style={draw=black!55, rounded corners=1.5pt, fill=white, minimum width=16mm, minimum height=5mm, inner sep=1pt},
    ordfl/.style={draw=wongsky!70!black, line width=1pt, fill=wongsky!35},
    durfl/.style={draw=wongorange!85!black, line width=1pt, fill=wongorange!30},
    cacheA/.style={draw=wongvermilion, densely dashed, line width=0.8pt, rounded corners=2pt, fill=wongvermilion!5, minimum width=24mm, minimum height=7.5mm},
    hostB/.style={draw=black!60, rounded corners=2pt, fill=black!5, minimum width=13mm, minimum height=6.5mm},
    dev/.style={draw=black!60, rounded corners=2pt, fill=white, minimum width=26mm, minimum height=9mm},
    on/.style={draw=black!85, fill=black!78, text=white, minimum width=8mm, minimum height=5.2mm, inner sep=0pt, font=\footnotesize\bfseries},
    off/.style={draw=black!45, densely dashed, fill=white, text=black!45, minimum width=8mm, minimum height=5.2mm, inner sep=0pt},
    ghost/.style={draw=wongvermilion, densely dotted, line width=0.8pt, fill=white, text=wongvermilion, minimum width=8mm, minimum height=4.6mm, inner sep=0pt, font=\footnotesize\bfseries},
    mv/.style={->, line width=1pt},
    lab/.style={font=\scriptsize, inner sep=1.5pt},
    head/.style={font=\footnotesize\bfseries, anchor=south},
    verdict/.style={font=\footnotesize\bfseries, anchor=north}]

  \node[head] at (1.55,4.55) {One operation of host A};
  \node[stepb] (s1) at (1.5,3.95) {store $x$};
  \node[stepb, ordfl] (f1) at (1.5,3.2) {flush $x$};
  \node[stepb] (s2) at (1.5,2.45) {store $y$};
  \node[stepb, durfl] (f2) at (1.5,1.7) {flush $y$};
  \node[lab, anchor=west, text=wongsky!55!black, font=\scriptsize\bfseries] at ([xshift=2pt]f1.east) {ordering};
  \node[lab, anchor=west, text=wongorange!75!black, font=\scriptsize\bfseries] at ([xshift=2pt]f2.east) {durability};
  \draw[->, black!60, line width=0.6pt] (s2.west) to[out=180, in=180, looseness=1.7] node[lab, left=1pt, text=black!65, align=right] {depends\\[-2pt]on} (s1.west);

  \draw[black!20, line width=0.6pt] (3.55,-0.1) -- (3.55,4.95);
  \draw[black!20, line width=0.6pt] (8.3,-0.1) -- (8.3,4.95);

  \newcommand{\podpanel}{%
    \node[cacheA] (A) at (0,3.9) {};
    \node[lab, anchor=south west, text=wongvermilion, font=\scriptsize\bfseries] at ([yshift=0.5pt]A.north west) {A crashes, cache lost};
    \node[dev] (D) at (0,2.3) {};
    \node[lab, anchor=north west, text=black!60] at ([xshift=-1pt]D.south west) {device};
    \node[hostB] (B) at (0.6,0.75) {B};
  }

  \node[head] at (5.9,4.55) {Without the \textcolor{wongsky!55!black}{ordering} flush};
  \begin{scope}[shift={(5.9,0)}]
    \podpanel
    \node[ghost] at (-0.6,3.9) {$x$};
    \node[off] (x) at (-0.6,2.3) {$x$};
    \node[on] (y) at (0.6,2.3) {$y$};
    \draw[mv] (0.6,3.52) -- node[lab, right] {write-back} (y.north);
    \draw[mv] (y.south) -- node[lab, right] {load} (B.north);
    \draw[->, wongvermilion, line width=1pt] (y.west) -- (x.east);
    \node[verdict, text=wongvermilion] at (0,0.3) {$\boldsymbol{\times}$\; B saw $y$; $x$ is gone};
  \end{scope}

  \node[head] at (12.85,4.55) {Without the \textcolor{wongorange!75!black}{durability} flush};
  \begin{scope}[shift={(10.6,0)}]
    \podpanel
    \node[on] (x) at (-0.6,2.3) {$x$};
    \node[on] (y) at (0.6,2.3) {$y$};
    \draw[mv, wongsky!60!black] (-0.6,3.52) -- node[lab, left, text=wongsky!45!black] {flush} (x.north);
    \draw[mv] (0.6,3.52) -- node[lab, right] {write-back} (y.north);
    \draw[mv] (y.south) -- node[lab, right] {load} (B.north);
    \draw[->, black!55, line width=0.7pt] (y.west) -- (x.east);
    \node[verdict, text=wonggreen!80!black] at (0,0.3) {$\checkmark$\; B saw $y$; both are there};
  \end{scope}
  \begin{scope}[shift={(15.1,0)}]
    \podpanel
    \node[ghost] at (0.6,3.9) {$y$};
    \node[on] (x) at (-0.6,2.3) {$x$};
    \node[off] (y) at (0.6,2.3) {$y$};
    \draw[mv, wongsky!60!black] (-0.6,3.52) -- node[lab, left, text=wongsky!45!black] {flush} (x.north);
    \node[lab, anchor=east, text=black!60] at ([xshift=-1pt]B.west) {idle};
    \node[verdict, text=wonggreen!80!black] at (0,0.3) {$\checkmark$\; no host saw $y$};
  \end{scope}

  \node[font=\small\bfseries, text=wongsky!45!black] at (5.9,-0.5) {required};
  \node[font=\small\bfseries, text=wongorange!70!black] at (12.85,-0.5) {not required};
\end{tikzpicture}

\caption{The claim. Host A stores $x$ and then $y$, which depends on $x$; host B may load $y$; then A crashes and its cache is lost. A black cell is a value on the device and a dashed cell one that is not. Without the \emph{ordering flush}, of $x$ before the store to $y$, B's load makes the fabric write $y$ back while $x$ is still only in A's cache: B has seen a value whose dependency the crash removes. Without the \emph{durability flush}, of $y$ after the operation, nothing that a survivor has seen is lost: if B has loaded $y$, the fabric has written it back first, and if no host has, $y$ disappears with A and no survivor can tell. A host crash requires the first flush and not the second.}
\Description{A diagram in three parts. Left: one operation of host A as four steps, store x, flush x (marked ordering), store y (which depends on x), flush y (marked durability). Middle, without the ordering flush: A's cache, lost in the crash, still holds x; the device holds y, written back because host B loaded it, and does not hold x; B saw y and x is gone; verdict, required. Right, without the durability flush, two cases: B loads y, and the device holds both x, flushed by A, and y, written back for B's load; or B is idle, and the device holds x while y is lost with A's cache and no host saw it; verdict, not required.}
\label{fig:claim}
\end{teaserfigure}
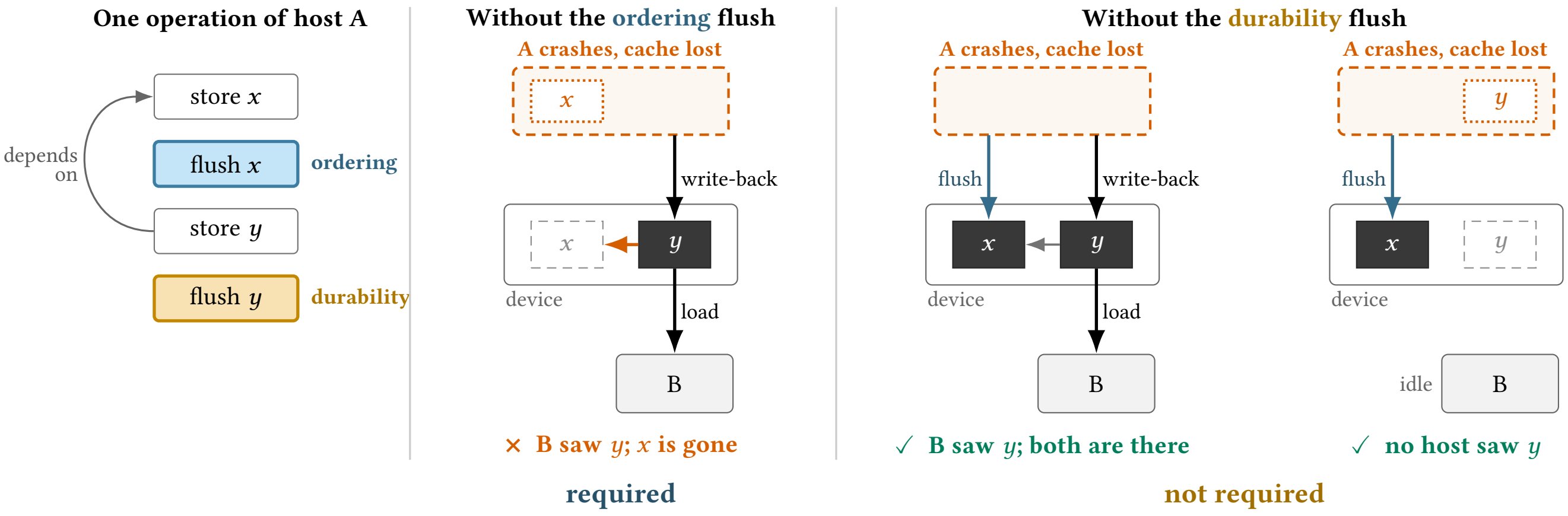

\maketitle

\section{Introduction}
\label{sec:intro}

Memory shared over Compute Express Link (CXL) lets a \emph{CXL pod}, a set of hosts attached to one memory device~\cite{lupin,tigon2025}, load and store the same cache lines through the hosts' own processor caches~\cite{dassharma2024}. We write \emph{pod} for short. A host that crashes takes its cache with it. The lines it has modified and not yet written back are gone, while the device and the other hosts go on.

Software that relies on coherence between hosts answers this loss in three ways (\S\ref{sec:related}). Some designs stop the whole pod when any host crashes. Some assume that stored energy or the platform flushes a crashing host's cache. The rest flush, and they flush as code for persistent memory does: an operation puts what it wrote on the device before it returns. That practice was shaped by the crash of a whole machine, after which no cache holds anything. A host crash in a pod differs in one respect. On a fabric that writes a modified line back to the device before it serves another host's access to that line, whatever a surviving host has seen of a line is already on the device. CXL's shared regions with hardware coherence are such a fabric on our reading of the specification (\S\ref{sec:model}). None of these designs says which of its flushes a host crash then requires.

\textbf{We show that a host crash requires the flushes that order one line before another, and none of the flushes that make a finished operation durable: by a proof for operations that take effect in a single store, and by tests on models of three published designs whose operations store several times.} A design for persistent memory issues both kinds (Figure~\ref{fig:claim}). An \emph{ordering flush} keeps a line from reaching the device before a line it depends on: a link before the record it names, a lock's release before the writes it covered. A \emph{durability flush} puts an operation's own stores on the device so that the operation survives. Durability flushes can go because a value they would protect is of one of two kinds. Either a surviving host has observed the value, in which case the fabric has already put it on the device, or only the crashed host has observed it, in which case, for a host whose clients die with it, the crash condition we use lets it vanish with that host.

\begin{figure*}[t]
\centering
\begin{tikzpicture}
\begin{axis}[paper, name=p0, scale only axis, width=32mm, height=17mm, xbar, bar width=5pt, xmin=0, xmax=165, ymin=0.4, ymax=4.6, xtick={0,50,100}, ytick={1,2,3,4}, yticklabels={{ordering flushes}, {ordering, one left out}, {the source's flushes}, {no flush}}, ytick style={draw=none}, xlabel={instances with a violation (\%)}, title={(a) Hash index}, title style={font=\footnotesize\bfseries, yshift=-4pt}]
\addplot[fill=wongvermilion, draw=none] coordinates {(55.56,4) (0.00,3) (22.22,2) (0.00,1)};
\node[anchor=west, font=\scriptsize] at (axis cs:57.56,4) {5 of 9};
\node[anchor=west, font=\scriptsize, text=wonggreen!70!black] at (axis cs:2.00,3) {0 of 9};
\node[anchor=west, font=\scriptsize] at (axis cs:24.22,2) {2 of 9};
\node[anchor=west, font=\scriptsize, text=wonggreen!70!black] at (axis cs:2.00,1) {0 of 9};
\end{axis}
\begin{axis}[paper, name=p1, at={(p0.east)}, anchor=west, xshift=7mm, scale only axis, width=32mm, height=17mm, xbar, bar width=5pt, xmin=0, xmax=165, ymin=0.4, ymax=4.6, xtick={0,50,100}, ytick={1,2,3,4}, yticklabels={{}, {}, {}, {}}, ytick style={draw=none}, xlabel={instances with a violation (\%)}, title={(b) Allocator, redo and era flushes kept}, title style={font=\footnotesize\bfseries, yshift=-4pt}]
\addplot[fill=wongvermilion, draw=none] coordinates {(92.86,4) (92.86,3) (75.00,2) (0.00,1)};
\node[anchor=west, font=\scriptsize] at (axis cs:94.86,4) {13 of 14};
\node[anchor=west, font=\scriptsize] at (axis cs:94.86,3) {13 of 14};
\node[anchor=west, font=\scriptsize] at (axis cs:77.00,2) {3 of 4};
\node[anchor=west, font=\scriptsize, text=wonggreen!70!black] at (axis cs:2.00,1) {0 of 14};
\end{axis}
\begin{axis}[paper, name=p2, at={(p1.east)}, anchor=west, xshift=37mm, scale only axis, width=32mm, height=17mm, xbar, bar width=5pt, xmin=0, xmax=165, ymin=0.4, ymax=4.6, xtick={0,50,100}, ytick={1,2,3,4}, yticklabels={{ordering flushes (both)}, {data flush only}, {lock flush only}, {no flush}}, ytick style={draw=none}, xlabel={instances with a violation (\%)}, title={(c) Lock table}, title style={font=\footnotesize\bfseries, yshift=-4pt}]
\addplot[fill=wongvermilion, draw=none] coordinates {(100.00,4) (100.00,3) (100.00,2) (0.00,1)};
\node[anchor=west, font=\scriptsize] at (axis cs:102.00,4) {9 of 9};
\node[anchor=west, font=\scriptsize] at (axis cs:102.00,3) {9 of 9};
\node[anchor=west, font=\scriptsize] at (axis cs:102.00,2) {9 of 9};
\node[anchor=west, font=\scriptsize, text=wonggreen!70!black] at (axis cs:2.00,1) {0 of 9};
\end{axis}
\end{tikzpicture}
\caption{Outcomes of the three studies. A bar is the share of explored instances in which some execution violates the design's crash condition when a host crashes and its cache is lost (for the lock table, the nine programs explored with up to two crashes, counting the program explored at two budgets once). With the ordering flushes alone, none does; leaving one of them out brings violations back. Every allocator bar keeps the source's flushes of redo entries and eras. Some runs stopped at a cap on the number of states; \S\ref{sec:locktable} to \S\ref{sec:allocator} give the counts.}
\Description{Three panels of horizontal bars, one per design, each with four flush disciplines. Hash index: no flush, violations in 5 of 9 instances; the source's flushes, 0 of 9; ordering flushes with one left out, 2 of 9; ordering flushes, 0 of 9. Allocator: no flush, 13 of 14; the source's flush, 13 of 14; one left out, 3 of 4; ordering flushes, 0 of 14. Lock table: no flush, 9 of 9; the lock flush only, 9 of 9; the data flush only, 9 of 9; both ordering flushes, 0 of 9.}
\label{fig:outcomes}
\end{figure*}
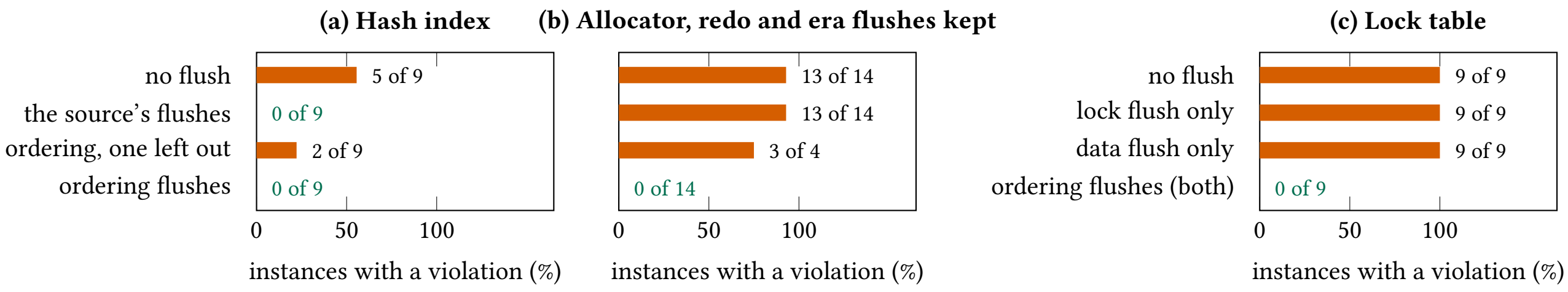

The contributions are these.
\begin{itemize}
  \item \emph{The ordering rule} (\S\ref{sec:ordering}). For one dependency between two lines, two rules on stores keep the device consistent through any crashes, and a dependent value stored while the device lacks the value it depends on can be exposed to a surviving host.
  \item \emph{A theorem for operations with one commit} (\S\ref{sec:ordering}). If every operation takes effect in one store and first flushes the other lines it read or prepared, an implementation that is correct without crashes meets our crash condition; an operation on one line needs no flush. If every finished operation must survive instead, even a register that one host writes alone needs a flush in every write.
  \item \emph{A lock table} (\S\ref{sec:locktable}) loses no lock of a surviving host at any sampled crash with no flush at all, and two ordering flushes keep every thread from reading part of a crashed host's update.
  \item \emph{A persistent hash index} (\S\ref{sec:hashindex}) with the flushes of two dependencies shows no violation for survivors in the explored instances, and its inserts write \HiOrdLinesLoad lines to the device each whereas those of its published code write \HiSrcLoad.
  \item \emph{An allocator} (\S\ref{sec:allocator}) flushes each new reference when it allocates. With that flush a survivor can hold or reach a freed object in every explored instance in which a reference or a segment leaves the client that crashes, and with the ordering flushes in its place in none.
\end{itemize}
The three designs are tested by exploring small instances of models of them on an executable model of the memory, exhaustively whenever the state space could be finished (Figure~\ref{fig:outcomes}; \S\ref{sec:method}, \S\ref{sec:limits}).

\textbf{For a designer the result is a rule of placement: flush a line before the store that depends on it, and nowhere else.} It replaces flushing everything the operation wrote, and it can ask for more flushes as well as fewer: one of the allocator's ordering flushes comes when a reference changes hands, at which point its published code flushes a redo entry and none of the lines the receiver will read. For hardware the result puts a price on one property of the fabric. When a modified line is instead passed from cache to cache, a crash loses a surviving host's lock in a share of the sampled crashes that depends on the transaction mix and the layout, and whose mean over seeds reaches \LkAllUseUp percent.

\section{Setting and model}
\label{sec:model}

A \emph{pod} is a set of hosts that load and store the lines of one memory device through their own processor caches~\cite{lupin,tigon2025}. The hosts crash independently of the device and of each other. We study the crash of a host: its cache is lost, and the device and the other hosts keep running.

\paragraph{States.}
The model, \mpod, restricts the operational model of Assa et al.~\cite{cxl0}, in its variant in which a load by another host writes the line back, to a pod: hosts own no shared memory, one device owns every shared line, and a line moves between hosts only through the device. It has hosts $1,\dots,n$, lines $x$ and values $v$. A state has a cache per host, which maps each line to nothing, to a clean copy $\clean(v)$ or to a dirty copy $\dirty(v)$, and a device memory, which maps each line to a value. A clean copy is one the host obtained from the device. A dirty copy is the value left by the host's last store or read-modify-write to the line and not yet written back. Each host also has private memory that no other host can access.

\paragraph{Steps.}
A thread on host $i$ takes the first four steps below; the fifth is a crash of the host, and the last two are silent steps of the hardware. A non-temporal store, which one of the designs we study uses, is a store followed by a flush of its line.
\begin{itemize}
  \item \textsc{Store}$_i(x,v)$. Allowed only when no other host holds $x$ dirty, this step drops $x$ from every other cache and leaves $\dirty(v)$ at $i$.
  \item \textsc{Load}$_i(x)$. If $i$ holds a copy, the load returns its value. Otherwise the load is allowed only when no host holds $x$ dirty; it returns the device value and installs a clean copy.
  \item \textsc{RMW}$_i(x,f)$, an atomic read-modify-write. Allowed only when no other host holds $x$ dirty, this step reads the value $r$ that a load would return, drops $x$ from every other cache, and leaves $\dirty(f(r))$ at $i$.
  \item \textsc{Flush}$_i(x)$. If $i$ holds $\dirty(v)$, the device memory of $x$ becomes $v$ and the copy becomes clean, and the thread waits for this. Otherwise nothing changes.
  \item \textsc{Crash}$_i$. A crash erases the cache and the private memory of $i$, leaving the device memory and the other caches untouched.
  \item \textsc{WB}$_i(x)$ (silent). If $i$ holds $\dirty(v)$, the device memory of $x$ becomes $v$ and the copy becomes clean.
  \item \textsc{Evict}$_i(x)$ (silent). A clean copy is dropped.
\end{itemize}
The silent steps are scheduled by an adversary and are invisible to threads: no thread can assume that one has occurred, and a dirty line can reach the device at any moment. A state is \emph{reachable} if a sequence of steps leads to it from empty caches and some initial device memory.

\paragraph{The premise.}
The conditions of \textsc{Store}, \textsc{Load} and \textsc{RMW} on other hosts' dirty copies say one thing about the hardware: \emph{an access to a line that another host holds modified is served only after that host has written the line back to the device.} A coherence protocol has this property or lacks it by design.

CXL's shared regions with multi-host hardware coherence have it, on our reading of revisions 3.0 and 4.0 of the specification~\cite{cxl30spec,cxl40spec}. For each shared region with back-invalidation the fabric manager chooses hardware coherence or coherence managed by software~\cite[\S2.4.4]{cxl30spec}; we consider the first. When the device needs a line back, as when another host requests it, it sends the holder a back-invalidate snoop. The holder's response carries no data: it reports the state of the holder's cache, and the response that reports the invalid state means that the holder has completed the snoop~\cite[Tables 3-37 and 3-38]{cxl30spec}. Data travels from a host to the device on one channel, that of requests with data, and no channel carries data from one host to another. A modified line can therefore leave its holder only by a write to the device, and the device serves the waiting host from its own memory. That the write completes before the response is the holder's obligation. The specification says so in its description of a snoop caused by an eviction from the device's snoop filter: the host finds that it holds modified data for the line, that data has to be written to the device before the snoop can complete, and the host sends its response once the device has completed the write~\cite[\S3.5.1.2 and Figure 3-20]{cxl30spec}. The statement belongs to a flow that the specification gives as an example; its summary of the rules for back-invalidation refers to these flows and has no separate rule for the order~\cite[\S3.3.3]{cxl30spec}. The order also follows from coherence: a holder that responded while its write was in flight would let the device serve a value that the write then replaces.

Two earlier works bear on this reading. Assa et al.\ write that the standard ``does not explicitly require remote loads to update memory'' and give the write-back as a variant of their model~\cite{cxl0}. Guo et al.\ adopt it as the memory model of their checker, in which ``remote reads ensure that cache lines have been written to the backing storage''~\cite{cxlmc}. Other coherent interconnects lack the property: the specifications of Arm's Coherent Hub Interface (CHI)~\cite{chispec} and of the Cache Coherent Interconnect for Accelerators (CCIX)~\cite[Tables 3-16 and 3-18]{ccixspec} let a snooped cache supply a modified line to the requester without a write to memory, and leave the duty to update memory with the holder or pass it to the requester. A fabric that hands the modified line from one cache to another is tested in \S\ref{sec:locktable}, in which switching the premise off makes survivors lose locks.

\begin{lemma}[coherence]
\label{lem:inv}
In every reachable state, if a host holds $x$ dirty then no other host holds $x$, and every clean copy of $x$ equals the device value of $x$.
\end{lemma}

\begin{lemma}[an observed value is on the device]
\label{lem:obs}
In a reachable state, if a load of $x$ by host $j$ returns $v$ and $j$ does not hold $x$ dirty, then the device value of $x$ is $v$.
\end{lemma}

\begin{proof}
Lemma~\ref{lem:inv} is by induction on the steps: a store or a read-modify-write leaves one holder; a load from the device needs no dirty holder and installs the device value; a flush or a write-back makes the dirty holder's value the device value; a crash and an eviction only remove copies. For Lemma~\ref{lem:obs}, a load by a host that does not hold the line dirty returns the host's clean copy, which equals the device value by Lemma~\ref{lem:inv}, or the device value itself.
\end{proof}

Lemmas~\ref{lem:inv} and~\ref{lem:obs} give the whole content of the premise for software: whatever a second host has seen of a line is on the device, and because no other host holds a line that $i$ holds dirty, a crash of $i$ loses only values that $i$ alone had.

\paragraph{What must survive a crash.}
A history records the invocations and responses of operations by threads, together with crash events. A crash ends an \emph{incarnation} of its host; threads the host runs afterwards belong to a new incarnation. Write $H^\circ$ for a history $H$ with its crash events removed. Durable linearizability (\dl) requires $H^\circ$ to be linearizable, so every operation that completed before a crash survives it, the crashed host's included~\cite{izraelevitz2016}; Assa et al.\ adopt it for partial crashes~\cite{cxl0}. We use a weaker condition.

\begin{definition}[observation-durable linearizability (\odl)]
\label{def:odl}
$H$ is observation-durably linearizable if for each crash event $c$ there is a set $S_c$ of operations, invoked by threads of the incarnation that $c$ ends and completed before $c$, such that $H^\circ$ without the operations in the sets $S_c$ is linearizable.
\end{definition}

A crash may remove completed operations of the host that crashed and no others. Every response a surviving host returned stays justified, because the linearization keeps every completed operation of a host that did not crash; so an operation of the crashed host whose effect a survivor's response depends on cannot be removed. \odl is the condition for a host whose clients die with it. A response that leaves the pod before the lines it depends on are on the device is outside its protection.

The set $S_c$ need not be the last operations of the crashed host. Let a host insert $a$ and then $b$ into two lines with no flush, let another host look up $b$, and let the first host crash. The device has $b$ and may not have $a$, and \odl accepts the history by removing the insert of $a$: a later operation of the crashed host survives and an earlier one does not. Buffered durable linearizability, the relaxation of \dl for crashes of a whole machine, also lets a crash remove completed operations, but only those outside a cut of the history that is closed under happens-before, so that what survives is ``consistent but not necessarily up to date''~\cite{izraelevitz2016}. \odl asks less of the crashed host's operations, which it does not keep in order, and more of the others: it removes no operation of a host that did not crash. A program that needs an operation of a host to survive whenever a later operation of the same host does has to order the two; \S\ref{sec:ordering} says what that takes.

\section{Ordering flushes and durability flushes}
\label{sec:ordering}

A design for persistent memory flushes for two reasons. An \emph{ordering flush} keeps one line from reaching the device before another line that it depends on. A \emph{durability flush} puts an operation's own stores on the device before the operation responds, so that the operation survives a crash. This section makes the first kind precise in \mpod, proves that it is sufficient for the dependency it protects and that a state it excludes can be exposed to a surviving host, proves that for operations that take effect in one store the ordering flushes give \odl, and shows why the second kind has a price that \odl does not pay.

\paragraph{Ordering pairs.}
Crash-consistent designs keep invariants of the form that if line $y$ holds a value of one kind, then line $x$ holds a value of another: a link names a record that has been initialized; a count covers the reference that raised it. An \emph{ordering pair} $(x,A,y,B)$ consists of two distinct lines $x$ and $y$ and two sets of values $A$ and $B$; we say that a $B$ value at $y$ \emph{depends on} an $A$ value at $x$. In a state,
\begin{itemize}
  \item $B$ is \emph{present} at $y$ if the device value of $y$ is in $B$ or some host holds a dirty copy of $y$ that is in $B$;
  \item $A$ is \emph{settled} at $x$ if the device value of $x$ is in $A$ and every dirty copy of $x$ is in $A$.
\end{itemize}
A state \emph{keeps the pair} if, whenever $B$ is present at $y$, $A$ is settled at $x$. Two rules constrain the steps that write, \textsc{Store} and \textsc{RMW}. Let such a step write the value $w$ to the line $z$ from the state $s$.
\begin{itemize}
  \item[\textbf{R1}] If $z=y$ and $w\in B$, then $A$ is settled at $x$ in $s$.
  \item[\textbf{R2}] If $z=x$ and $B$ is present at $y$ in $s$, then $w\in A$.
\end{itemize}
R1 says that the dependent value is written only once the value it depends on is on the device. R2 says that the value depended on is not withdrawn while a dependent value exists.

\begin{theorem}[the rules keep the order]
\label{thm:kept}
Let a state keep the pair $(x,A,y,B)$, and let every \textsc{Store} and \textsc{RMW} step of a run from that state obey R1 and R2. Then every state of the run keeps the pair, whatever crashes, write-backs and evictions the run contains. In particular, in every state of the run, if the device value of $y$ is in $B$ then the device value of $x$ is in $A$.
\end{theorem}

A state with empty caches whose device value of $y$ is not in $B$ keeps the pair, so the theorem covers every run from such an initial state.

\begin{proof}
We show that each step preserves the property. Let the state before the step keep the pair.

\emph{A store or read-modify-write of $w$ to $y$ by host $i$.} Afterwards the only dirty copy of $y$ is $w$, at $i$, and the device is unchanged. If $w\in B$, R1 gives that $A$ was settled at $x$, and the step touches no copy of $x$. If $w\notin B$ and $B$ is present at $y$ afterwards, the device value of $y$ is in $B$, so $B$ was present before, $A$ was settled at $x$, and still is.

\emph{A store or read-modify-write of $w$ to $x$ by host $i$.} Afterwards the only dirty copy of $x$ is $w$, and the device and every copy of $y$ are unchanged. If $B$ is present at $y$, it was present before, so the device value of $x$ is in $A$, and R2 gives $w\in A$.

\emph{A flush or write-back of a dirty copy $v$ by host $i$.} If the line is $y$, the device value of $y$ becomes $v$ and the copy becomes clean; if $B$ is present at $y$ afterwards, then $v\in B$ or another dirty copy is in $B$, so $B$ was present before, and $x$ is untouched. If the line is $x$, the device value of $x$ becomes $v$ and the presence of $B$ at $y$ does not change; if $B$ is present, $A$ was settled, so $v\in A$ and the remaining dirty copies are in $A$.

\emph{A load or an eviction} changes no device value and no dirty copy. \emph{A crash} removes copies and changes no device value: it cannot make $B$ present at $y$, and it cannot unsettle $A$ at $x$. A step on a third line changes nothing at $x$ or $y$.

For the last sentence, a device value of $y$ in $B$ makes $B$ present, so $A$ is settled, and the device value of $x$ is in $A$.
\end{proof}

\begin{proposition}[a store out of order can be exposed]
\label{thm:needed}
In a reachable state, let host $i$ hold a dirty copy $v\in B$ of $y$ while the device value of $x$ is not in $A$, and let $j\neq i$. Then the steps \textsc{WB}$_i(y)$, \textsc{Load}$_j(y)$ and \textsc{Crash}$_i$ can occur in this order; the load returns $v$; and afterwards the device holds $v\in B$ at $y$ and a value outside $A$ at $x$.
\end{proposition}

\begin{proof}
By Lemma~\ref{lem:inv} no host other than $i$ holds $y$. The write-back puts $v$ on the device and leaves no dirty copy of $y$, so the load by $j$ is served from the device and returns $v$. None of the three steps changes the device value of $x$.
\end{proof}

The write-back in Proposition~\ref{thm:needed} is the one the premise performs before it serves $j$; the adversary adds only the crash. If $i$ had stored an $A$ value to $x$ and not flushed it, the crash erased it: a surviving host has read a $B$ value, and the $A$ value it depends on exists nowhere. The proposition concerns the device's value of $x$. R1 also asks that no host hold a dirty copy of $x$ outside $A$, for a plainer reason: a silent write-back of such a copy puts it on the device at any moment.

\paragraph{What the theorems say about flushes.}
Consider a host that has stored an $A$ value to $x$ and is about to store a $B$ value to $y$. R1 asks that the device hold an $A$ value at $x$. The host cannot see silent write-backs, so it can know this in two ways: it has flushed $x$ since its last store to $x$, or another host has accessed $x$ since (Lemma~\ref{lem:obs}), which the host knows only if the program tells it. Either way the program has to keep other hosts from storing a value outside $A$ to $x$ in between, by a lock or by giving the line one writer; R2 forbids such a store only once a $B$ value is present. The flush of $x$ before the store to $y$ is the ordering flush of the pair, and Proposition~\ref{thm:needed} shows what its absence allows. Theorem~\ref{thm:kept} holds for each pair of a design separately, so a design with several pairs keeps them all by obeying the two rules for each. Two values on one line need no pair: a line reaches the device whole.

Theorem~\ref{thm:kept} never asks for a flush of $y$. The pair is kept whether or not the dependent value ever reaches the device, and if the storing host crashes first, the value is lost together with everything that depended on it in that host's cache. \textbf{This is the sense in which a host crash requires a design's ordering flushes and none of its durability flushes:} what a durability flush would protect is either something a surviving host has observed, which Lemma~\ref{lem:obs} already puts on the device, or something only the crashed host had, which \odl lets a crash remove. The argument is about one line. An operation that wrote several lines can be observed through one of them, and the others must then be ordered before that one by pairs of their own. The same holds for two operations of one host that must survive in order (\S\ref{sec:model}): the later store depends on the earlier, and the earlier line is flushed first.

\paragraph{From dependencies to the crash condition.}
Theorem~\ref{thm:kept} and Proposition~\ref{thm:needed} concern one pair. For operations of one common form the pairs can be read off the program, and keeping them gives \odl. An \emph{implementation} of an object gives each operation a program: loads, stores, read-modify-writes and flushes taken by one thread, each determined by the operation's arguments and by the values that its earlier loads and read-modify-writes returned, and then a response determined in the same way. It is \emph{linearizable without crashes} if the history of every execution that contains no crash is linearizable~\cite{herlihywing1990}. An object may restrict how it is called, as a stack does that takes a new node with each push; its executions are then those that obey the restriction. The proof of Theorem~\ref{thm:commit} leaves operations of crashed hosts out of an execution, and we assume that what remains still obeys the restriction. A restriction on each call by itself, as the stack's is, survives the removal of any operations.

\begin{lemma}[what a host alone has seen]
\label{lem:alone}
In a run from a reachable state, let host $i$ hold line $x$ dirty, and consider the steps that follow, up to and not including the first flush or write-back of $x$ by $i$ or crash of $i$. Throughout them $i$ holds $x$ dirty, no other host loads, stores, read-modify-writes, writes back or evicts $x$, and the device value of $x$ does not change. If the step that comes next is a crash of $i$, the device value of $x$ is the same after the crash, and no host holds $x$.
\end{lemma}

\begin{proof}
While $i$ holds $x$ dirty, a store, a read-modify-write or a load from the device of $x$ by another host is not allowed, and by Lemma~\ref{lem:inv} no other host holds a copy of $x$ to load, write back or evict. A step of $i$ other than a flush, a write-back or a crash leaves its copy of $x$ dirty. The device value of $x$ changes only by a flush or a write-back of $x$ by the host that holds it dirty. The crash erases the only copy.
\end{proof}

Three terms say how a value travels through an execution. A \emph{write} is a store, or a read-modify-write that changes the value of its line. Let every value carry a mark that no program sees: a write marks the value it puts in its line with its own step, and every other step passes the mark on with the value. A read-modify-write that writes back the value it read is an \emph{attempt}, as a failed compare-and-swap is: it leaves the line dirty at its host and leaves the mark as it was. A load or a read-modify-write \emph{reads from} the write whose mark the value it returns carries. A write \emph{descends from} a write $C$ if it is $C$, or if its operation has read, at that step or an earlier one, from a write that descends from $C$.

An implementation has \emph{single commits} if in every execution, with or without crashes, the lines that each operation accesses divide into lines that it \emph{prepares} and others, as follows. On the others the operation takes loads and attempts and then at most one write, its \emph{commit}, and the commit is its last load, store or read-modify-write of any line. A line that an operation $q$ prepares belongs to $q$ until its commit publishes it: at any point of the execution, a step of another operation $r$ loads, stores or read-modify-writes the line only after $r$ has read from a write that descends from the commit of $q$. A linked stack whose pushes are each given a node that has a line to itself, and that no other push is given, has single commits. A push loads the head, fills its node, and swings the head to the node by a compare-and-swap; a swap that fails is an attempt, and the push starts again. A pop loads the head and the node that the head names, and swings the head to the next node in the same way. The address of a node is written only by the commit of its push and by operations that have loaded it, so no operation reaches the node in another way. The \emph{commit rule} is this: before its commit, an operation flushes every other line that it has loaded, stored or read-modify-written, after its last such step on that line.

\begin{theorem}[single commits]
\label{thm:commit}
Let an implementation with single commits be linearizable without crashes, and let its operations obey the commit rule. Then every execution satisfies \odl.
\end{theorem}

\begin{proof}
Call the \emph{current value} of a line the dirty copy of it, if a host holds one, and its device value otherwise. By Lemma~\ref{lem:inv} a load or a read-modify-write returns the current value of its line. A store or a read-modify-write sets the current value; a flush, a write-back and an eviction leave it as it is; a crash of host $i$ resets it to the device value for the lines that $i$ holds dirty and leaves it for the others. In an execution with no crash, therefore, a load returns the last value written to its line, as on a memory with no caches.

Fix an execution, with history $H$. For a crash $c$ of host $i$, call a line \emph{lost at $c$} if $i$ holds it dirty just before $c$, and let its \emph{run} extend from the last step at which $i$ came to hold it dirty up to $c$. Just before the run no host holds the line dirty, so its device value is its current value, and by Lemma~\ref{lem:alone} the device value does not change during the run. The crash therefore returns the line to the value and the mark that it had before the run: it \emph{undoes} the writes to the line during the run, and no later step reads from them. By induction on the steps, the current value of a line is at every step that of the last write to it that no crash has undone. Let $R_c$ be the set of operations that load, store or read-modify-write a line lost at $c$ during its run. By Lemma~\ref{lem:alone} they are operations of threads of $i$ in the incarnation that $c$ ends. Let $R$ be the union of the sets $R_c$.

\emph{A commit of an operation of $R_c$ is a write that $c$ undoes.} Let the operation access the lost line $x$ during the run of $x$ and commit on the line $y$. The commit is its last access, and it comes before $c$. If $y=x$, the commit is in the run. If $y\neq x$, the commit rule puts a flush of $x$ by the operation between that access and the commit; but $i$ holds $x$ dirty throughout the run, so the flush would write $x$ back, and $i$ would have to come to hold $x$ dirty again later, which contradicts the choice of the run. So each write of an operation of $R_c$ is its commit, which $c$ undoes, or a write to a line that the operation prepared.

\emph{No operation outside $R$ reads from a write of an operation of $R$.} Suppose that a step $t$ of an operation $r$ reads from a write $w$ of another operation $q$, that $q$ is in $R_c$, and that the claim holds for the steps before $t$. If $w$ is the commit of $q$, then $c$ undoes it, so $t$ comes before $c$, in the run of the line, and $r$ is in $R_c$. Otherwise $w$ is a write to a line that $q$ prepared, so $r$ has read, before $t$, from a write that descends from the commit of $q$. Every write before $t$ that descends from that commit is by an operation of $R$: the commit is, and any other is by an operation that had read from an earlier one, which puts the operation in $R$ by the claim for the steps before $t$. So $r$ read from a write of an operation of $R$ before $t$, and $r$ is in $R$.

\emph{Without the crashes and the operations of $R$, an execution remains.} Take the invocations, the steps and the responses of the operations outside $R$, in their order, and run them in \mpod from the same initial device memory, with a write-back before each step that needs one and with no crash and no other silent step. Each host runs the threads of all its incarnations, and a thread whose operation a crash interrupted takes no further step. By assumption the calls obey the object's restriction, if it has one. We show that each load and each read-modify-write returns the value that it returned in the original execution, so that each operation takes the same steps and gives the same response. In the original execution, let a step $t$ of an operation outside $R$ read from the write $w$ to the line $x$. The operation of $w$ is outside $R$. Every write to $x$ between $w$ and $t$ was undone before $t$, so it lies in a run of $x$, and its operation is in $R$. Hence $w$ is the last write to $x$ before $t$ by an operation outside $R$, and its value is what $t$ returns in the execution without crashes. If $t$ returns the initial value of $x$ instead, the same holds of every write to $x$ before $t$.

The execution without crashes has a linearizable history, and that history is $H^\circ$ without the operations of $R$. For each crash $c$ take as $S_c$ the operations of $R_c$ that completed; threads of the incarnation that $c$ ends invoked them, and they completed before $c$. The other operations of $R$ never respond, and a linearization may leave them out. Hence $H$ is observation-durably linearizable.
\end{proof}

The commit rule issues the ordering flushes of an operation's pairs: the lines it read or prepared, each paired with the line of its commit. It does not ask, as R1 does, that at the commit the device and every dirty copy of such a line hold a value of the kind depended on; linearizability without crashes already accounts for a line that another operation changes in between. It asks for no flush of the commit line, which is the absence of a durability flush. When every operation accesses one line only, by loads, attempts and at most one write last, no other line is read or prepared, so an implementation that is linearizable without crashes satisfies \odl with no flush by Theorem~\ref{thm:commit}. An acquisition of one lock by a read-modify-write on the line of its record, which leaves the record as it was if another thread holds the lock and may be retried, and a release by one store or read-modify-write take that one-line form, so a table of such locks keeps every lock that a surviving host took, with no flush (\S\ref{sec:locktable}). The restriction that only its holder releases a lock survives the removal in the proof: a release follows its acquisition on the same line, so a crash that undoes the acquisition undoes the release with it.

\paragraph{What is not proved.}
Theorem~\ref{thm:commit} does not cover an operation that writes more than once to lines it did not prepare, such as one that takes a lock, writes and releases it. A crash can then leave some of the operation's writes on the device and undo others, and which operations may be dropped is no longer decided by which lines were lost. Nor does it cover a design that gives a prepared line to a second operation, as one that recycles nodes or packs two nodes into a line does, or a program that keeps state in its host's private memory from one operation to the next or runs recovery code after a restart. The update paths of the three designs we study write several times. For them we identify the pairs by reading the design, or for the allocator by exploration (\S\ref{sec:allocator}), and we test by exploration that keeping the pairs gives the crash condition. The commit rule is also more than a design may need: it flushes every line an operation read, whereas a design can know that the line is on the device already, as a pop of the stack above knows of a node that its push flushed before publishing it.

\paragraph{Durability has a price.}
The weaker condition is what removes the durability flushes. Under \dl even a register needs one in every write.

\begin{proposition}[one flush per update under \dl]
\label{prop:lower}
Let a deterministic implementation of a read-write register satisfy \dl, and let each of its operations complete when run alone from any reachable state. Consider an execution in which one host runs one thread that performs writes of values distinct from each other and from the initial value, no other host takes a step, and no silent write-back occurs. Every write that completes in it contains a flush.
\end{proposition}

\begin{proof}
Suppose a write $u$ by host $i$ completes with no flush between its invocation and its response. No other host took a step and no silent write-back occurred, so the device memory at the response is as it was at $u$'s invocation. Crash $i$ right after the response, and let a thread of another host run a read alone. The read starts from the device memory and empty caches that it would start from in the execution in which $u$ was never invoked and $i$ crashed at $u$'s invocation, so it takes the same steps and returns the same value in both. In that execution the history is linearizable only if the read returns the value of the last write before $u$, or the initial value, and neither is $u$'s value. So the read returns a value other than $u$'s after $u$ completed, with no write in between, and the history with its crash removed is not linearizable.
\end{proof}

The proposition is the pod counterpart of the bound of one fence per update that Cohen, Guerraoui and Zablotchi prove for crashes of a whole system~\cite{cohen2018}. Under \odl the same execution is harmless: the crash may remove $u$, which no other host observed.

\section{How the designs are tested}
\label{sec:method}

No processor, switch or device available to us keeps caches coherent across hosts, and the testbeds in the literature, virtual machines on one server, cannot lose one host's cache. The three studies that follow therefore run on an executable form of \mpod, and they share one method.

\paragraph{A design as steps.}
We write the operations of a published design as sequences of the steps of \S\ref{sec:model}, one step for each access to shared memory that its published source code makes on the path studied: every load, store, compare-and-swap and flush, in the source's order, except when a study says that accesses were merged or left out. A \emph{discipline} is the design with a chosen set of flushes. The disciplines are: no flush; the flushes of the source, when the source has any; the \emph{ordering flushes}, which are the flushes of the ordering pairs we identify in the design and no others; and the ordering flushes with one of them left out (in the hash index, the flush after a remove; in the allocator, the line of an object that the sent object refers to). One more discipline, with a known error or with the premise switched off, must be caught, as a check that the search can find a violation. A model may also contain steps that its design needs on a pod and that its source lacks; the hash index has two, a lock that records its holder and one flush when a host starts, and \S\ref{sec:hashindex} describes them.

\paragraph{Exploration.}
An instance fixes two or three hosts, their threads, and a handful of operations per thread. The explorer visits every reachable state of the instance: every interleaving of the threads' steps, with a crash at any point, of any host in the lock-table study and of one designated host in the other two, within a budget of one or two crashes. It does not interleave the silent steps. At a crash it leaves each line that the crashed host held dirty with any one of the values the line has held dirty since it was last clean, or with the device's value. Each choice is what one schedule of silent write-backs leaves. Every schedule leaves one of them, because a write-back that no other host's access requires changes nothing a thread can observe until the host that held the line crashes. The design's crash condition is evaluated during every execution: after each step in the explorers of the lock table and the allocator, and in that of the hash index at the end of the execution and at each crash that leads to a state not visited before, so that its counts of checked histories depend on the order of the search. A run is \emph{exhausted} when no state remains to visit. A run that reaches its cap on the number of states first is reported as stopped, and a zero from a stopped run is a statement about the states it visited.

\paragraph{Controls.}
Every instance is also explored with no crash, under which no discipline may violate the condition; the exceptions are \HiNoCrashMissing instances of the hash index, which a first pass did not exhaust without a crash (it found no violation) and which were not run again. When the paper that describes a design states an assumption under which the design needs no flush, the instance is explored under that assumption as well.

\paragraph{What this method can show.}
A violation is an execution of our model of the design, with its steps listed, and can be checked against the source by hand. A zero from an exhausted run covers one instance.

\section{A lock table}
\label{sec:locktable}

\begin{figure*}[t]
\centering
\begin{tikzpicture}
\begin{axis}[paper, name=share, scale only axis, width=62mm, height=40mm,
    xmin=-3, xmax=85, ymin=0.4, ymax=10.6, xtick={0,20,40,60,80},
    ytick={1,...,10},
    yticklabels={{skew 0.99, 50\% reads, all cross-host}, {skew 0.99, 95\% reads, all cross-host}, {skew 0.99, 50\% reads, 10\% cross-host}, {skew 0.99, 95\% reads, 10\% cross-host}, {skew 0.7, 50\% reads, all cross-host}, {skew 0.7, 95\% reads, all cross-host}, {skew 0.7, 50\% reads, 10\% cross-host}, {skew 0.7, 95\% reads, 10\% cross-host}, {TPC-C, 58/90\% remote}, {TPC-C, 10/15\% remote}},
    ymajorgrids, grid style={black!12}, ytick style={draw=none},
    xlabel={sampled crashes that lose a survivor's lock (\%)},
    title={(a) Under pass-dirty forwarding}, title style={font=\footnotesize\bfseries, yshift=-4pt},
    legend style={at={(0.99,0.985)}, anchor=north east, font=\scriptsize, draw=none, fill=none}, legend cell align=left]
\addplot[only marks, wongorange, mark=*, mark size=2pt, mark options={solid, fill=wongorange}, error bars/.cd, x dir=both, x explicit] table[x=use, y=row, x error plus=usep, x error minus=usem] {data/locks_share.dat};
\addplot[only marks, wonggreen, mark=diamond, mark size=2.8pt, line width=0.9pt, error bars/.cd, x dir=both, x explicit] table[x=rand, y=row, x error plus=randp, x error minus=randm] {data/locks_share.dat};
\addplot[only marks, wongsky, mark=square*, mark size=1.5pt, mark options={solid, fill=wongsky}, error bars/.cd, x dir=both, x explicit] table[x=one, y=row, x error plus=onep, x error minus=onem] {data/locks_share.dat};
\legend{8 per line{,} in order of creation, 8 per line{,} at random, 1 per line}
\end{axis}
\begin{axis}[paper, at={(share.east)}, anchor=west, xshift=9mm, scale only axis, width=62mm, height=40mm,
    xbar stacked, bar width=5pt,
    xmin=0, xmax=34, ymin=0.4, ymax=10.6, ytick={1,...,10}, yticklabels={}, ytick style={draw=none},
    xlabel={stores per committed transaction},
    title={(b) Stores to lock records}, title style={font=\footnotesize\bfseries, yshift=-4pt},
    legend style={at={(0.98,0.60)}, anchor=east, font=\scriptsize, draw=none, fill=none}, legend cell align=left,
    legend image code/.code={\fill[#1] (0cm,-0.08cm) rectangle (0.28cm,0.08cm);}]
\addplot[xbar, fill=wongvermilion, draw=none] table[x=excl, y=row] {data/locks_flush.dat};
\addplot[xbar, fill=black!22, draw=none] table[x=other, y=row] {data/locks_flush.dat};
\legend{exclusive acquisitions, every other store}
\end{axis}
\end{tikzpicture}
\caption{A lock table on the executable model: the lock requests of two transaction mixes, TPC-C and a key-value mix with Zipfian keys, on 8 hosts with 3 workers each. (a) The share of sampled crashes in which, under pass-dirty forwarding (a fabric that hands a modified line from one cache to another without writing the device), at least one lock held by a transaction on a surviving host is lost, for three layouts of the 8-byte lock records; means over ten seeds with 95 percent $t$ intervals, cut at zero. The marks for eight records per line placed at random lie on those for one record per line, which they exceed by at most \LkRandGapMaxPts points. Without forwarding the share is zero in every run. (b) Stores to lock records per committed transaction, with records created at their first use by a second host. A design that flushes after every store issues the whole bar; the first ordering flush follows an exclusive acquisition, the darker part.}
\Description{Two panels over the same ten rows, one per transaction mix. Panel (a) is a dot plot of the share of sampled crashes in which a survivor's lock is lost under forwarding: with one record per line it is below 0.1 percent for the two TPC-C mixes, at most 2.9 percent at skew 0.7, and 8.5 to 52.2 percent at skew 0.99; with eight records per line placed at random it exceeds those values by at most 1.4 points; with eight per line in order of creation it is about 20 percent for TPC-C with 58/90 percent remote and 67 to 78 percent for the mixes whose transactions all cross hosts. Panel (b) is a stacked bar chart of stores to lock records per committed transaction, from under 1 to about 33, in which exclusive acquisitions are the darker part of each bar, between 2 and 37 percent of it.}
\label{fig:locks}
\end{figure*}
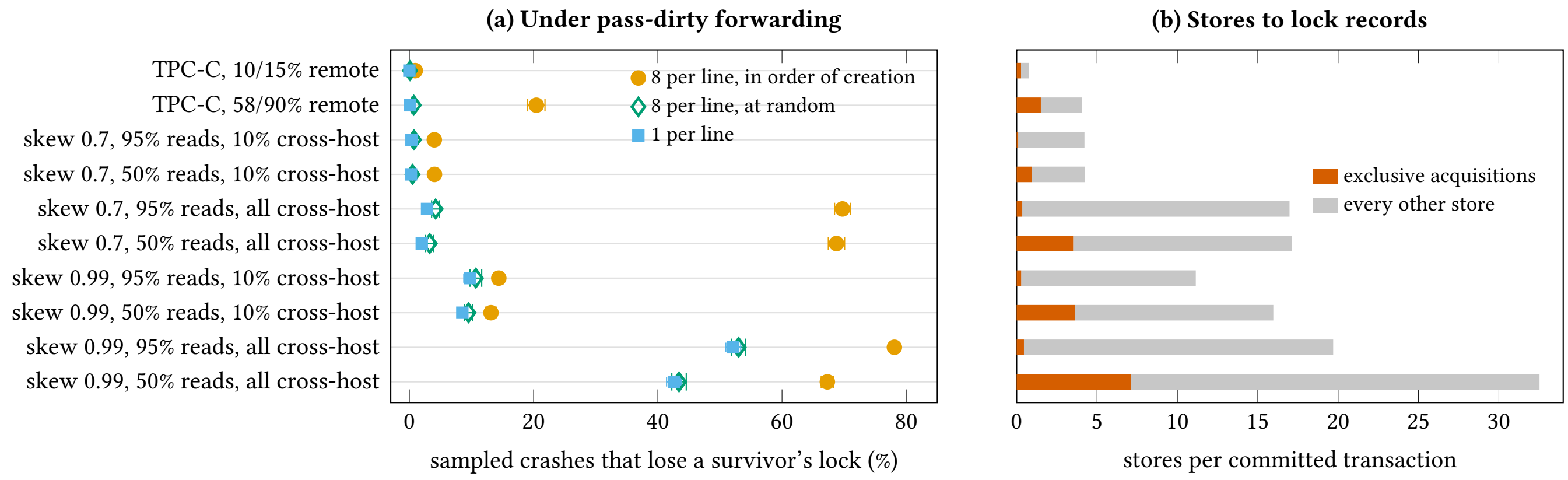

The databases designed for pods keep their cross-host locks under hardware coherence, when a device offers it, and the data that the locks guard outside it: Tigon keeps a latch and lock bits in an 8-byte record for each shared tuple~\cite{tigon2025}, and XTRA keeps a lock bit and a version in each slot of a directory~\cite{xtra2026}. Tigon treats the failure of any component as a failure of the whole system. XTRA has the surviving hosts scan the directory for the locks that a crashed host held and, from its redo log, roll each of its transactions forward or release its locks. We drove the executable model of \S\ref{sec:method} with the lock requests of two transaction mixes under two-phase locking, at settings taken from Tigon's evaluation, and at \LkCrashes sampled instants in \LkRuns runs compared the locks in force with the lock table that would remain if one host crashed at that instant. Without forwarding, no sampled crash loses a lock that a transaction on another host holds or brings back one that it released, as Lemma~\ref{lem:obs} requires and as Theorem~\ref{thm:commit} proves for locks taken and released on one line by one read-modify-write each. With \emph{pass-dirty forwarding}, in which an access to a line that another host holds modified is served from that host's cache and the device is not written, the share of sampled crashes that lose at least one such lock reaches a mean over seeds of \LkHotRdAllOne percent in one mix when each lock record has a line of its own, and of \LkHotRdAllUse percent when eight records share a line in the order of their creation (Figure~\ref{fig:locks}a).

The crashed host's own locks are another matter. Without forwarding, in every mix and layout at least \LkAbsentMinPct percent of the locks that its transactions hold in lock records are lost with it: no other host has accessed their lines since they were taken, and no eviction has written them back. \odl permits that loss, and it is harmless only if what the lock guards is lost with it. Nothing orders a host's lines on their way to the device, so a write under the lock can reach the device without the lock, and a release without the write. Two rules exclude both: a host flushes the lock's line after an exclusive acquisition, before its first write under the lock, and has its writes on the device before it releases. In an exhaustive exploration of nine small programs without forwarding (\LkExInstances instances: one program is explored with one crash and again with two), no execution under both rules lets a thread that is granted the lock read part of a crashed host's update, and with either rule dropped every instance has such an execution. The first rule costs from \LkExclMin to \LkExclMax flush steps per committed transaction on these mixes, whereas a flush after every store to a lock record issues from \LkStoresMin to \LkStoresMax (Figure~\ref{fig:locks}b). These are flush steps alone; \S\ref{sec:cost} adds the waits that unflushed lines cause. The second is the flush by which a commit publishes its updates before it releases its locks, as XTRA's does. Under both rules the lock of a host that crashed in mid-update stays in the lock table, over data that may be half written. A recovery that scans the lock table, as XTRA's does, can rely on finding every lock of the crashed host that another host observed, and one that no host observed only if its holder flushed it; before it releases such a lock it has to roll the update forward or back, for which XTRA flushes a redo log before the update's first write.

The two rules are the ordering flushes of two pairs in the sense of \S\ref{sec:ordering}: a value stored under the lock depends on the lock word that names its writer, and the lock word that frees the lock depends on the values stored under it. Neither is a durability flush: no shared acquisition, refused request or release is followed by a flush, and the lock table as a whole is never flushed. The result is the same with two crashes and in the three shared layouts we explored. In \LkLayoutBothRuns further exhaustive runs no execution violates the property under both flushes: the same instances with two crashes allowed (\LkLayoutSepRuns runs, two of them of the same program), and instances in which lock words share a line with a guarded value, with each other, or with both, with two crashes (\LkLayoutSharedTwoRuns runs) and with one (\LkLayoutSharedOneRuns). Every one of these runs has a violating execution under no flush and under either flush alone, and so has each of the \LkLayoutFwdRuns runs made with forwarding.

\section{A hash index}
\label{sec:hashindex}

P-CLHT is the persistent form, from RECIPE~\cite{recipe2019}, of a hash table whose buckets are one cache line each~\cite{clht2015}. A bucket holds a lock, three slots for key-value pairs and a link to the next bucket of its chain. A lookup takes no lock. An insert locks the head of the chain and puts the pair in a free slot; if the chain has none, it fills a new bucket and links it to the last one. A remove clears the key of the pair's slot. RECIPE makes each of these stores a non-temporal store or follows it with a flush, and adds a fence, so that an operation is on the device when it returns.

We wrote the three operations as steps from the source and explored them on two or three hosts with one crash, \HiInstances instances in all (\HiRuns runs, \HiStatesM million states, \HiNoCrashRuns of the runs with no crash). The table exists when an instance starts, and no instance resizes it. The source takes its buckets from the system's allocator and does not say how they are recovered; the model adds a record, on a line of its own, that a bucket is in use, and the model's form of the source flushes that record as well. The model's buckets have two slots. The crash condition is \odl; we also evaluate \dl.

Three things in the model are not in the source. The source's lock is a plain word, and the source frees every lock in one routine, which cannot run while other hosts are using the table. The model's lock therefore records the host and the incarnation of its holder, and a thread that finds the lock of an incarnation that has since restarted takes it over; Guo et al.\ change the locks of another of RECIPE's indexes in this way~\cite{cxlmc}. To number its incarnations a host stores and flushes one line when it starts, under every discipline, the one with no flush included. And the memory that a new bucket is taken from holds an old pair for the key being inserted, as after an earlier use, because the zeroing of a fresh bucket is itself a cached store that a crash can lose. A survivor's insert or remove that finds the lock held on the device by a host that has crashed and not restarted waits, and its operation stays pending in the history; the crashed host restarts in two of the instances.

\paragraph{Two ordering pairs.}
The first pair is the one every linked structure has: a link depends on the content of the bucket it names, and on the allocator's record that the bucket is in use. The second is particular to a table whose chains span lines: a key stored in a slot of one line depends on the removal of that key's old pair from another line. The ordering flushes are therefore two: the new bucket and its allocation record before the link, and the line of a removed pair before the key is stored again. Our discipline issues the second after every remove, which is more often than the pair requires. We fixed the two pairs, and predicted that the first alone would not be enough, before any exploration ran.

\paragraph{Results.}
With the flushes of the source, no history violates \odl or \dl (\HiPmChecks histories checked; \HiPmStopped of the \HiInstances explorations stopped at \HiCapM million states). With no flush, an insert that extends a chain can be lost while its link survives: in one exhausted instance \HiNoneOdlExt histories violate \odl, the first of them a lookup that follows the link to a bucket whose content never reached the device and returns the stale pair that the bucket's memory held before. With the first ordering flush alone such histories are gone, and \HiFirstPairOdl violating histories remain in another exhausted instance: a host removes a key and starts to insert it again with a new value, in a free slot of another line, and crashes during the insert, with the new pair's line on the device and the removal not; a survivor's remove of the key then returns the new value, and its next lookup returns the old one. \textbf{With both ordering flushes no history violates \odl} in any of the \HiInstances instances (\HiDepChecks histories checked; \HiDepExhausted state spaces exhausted, \HiDepStopped stopped at \HiCapM million states). The discipline with a known error stores the link before it fills the bucket and issues no flush; it is caught in \HiBadlinkInstances instances (\HiBadlinkMin to \HiBadlinkMax histories). In \HiBadlinkSame of them no insert extends a chain, the erroneous step never runs, and the counts are those of no flush. The control shows that the search finds violations; it does not separate the misplaced link from the missing flushes.

An insert into a free slot of the chain's first bucket, of a key that has no unflushed removal on another line, needs no flush at all: it touches one line, and under every discipline, the one with no flush included, no history of the instance that isolates it violates \odl. If the insert is lost with its host, \odl removes it; if a survivor has seen it, it is on the device (Lemma~\ref{lem:obs}).

The first pair also protects liveness. Without the flush of the allocation record, a host that restarts after its crash can take from its pool a bucket that is already linked, fill it and link it to itself, and a lookup then walks that bucket without end (\HiNeverNone checked histories with an operation that never returns, in the first \HiCapM million states of one instance under no flush; none under the first ordering flush).

\paragraph{What the durability flushes bought.}
Under the ordering flushes \dl fails in \HiDepDl of the checked histories, as Proposition~\ref{prop:lower} leads one to expect: \odl lets completed operations of the crashed host that no survivor observed be lost, and \dl does not. We counted on P-CLHT's own code, with counters added, on one thread, over \HiCountSeeds seeds of \HiCountOps operations per phase. Of the inserts into an empty table, \HiInPlacePct percent go to a free slot of an existing bucket. Per insert in that phase the source writes \HiSrcLoad lines to the device, by a flush or a non-temporal store (\HiSrcLoadResize with its \HiResizes resizes, which the model does not have), and the ordering flushes write \HiOrdLinesLoad, in \HiOrdLoad flush steps that each name a new bucket and its allocation record. In a phase of half inserts and half removes the source writes \HiSrcSteady lines per operation and the ordering flushes \HiOrdLinesSteady, \HiOrdSteadyRemove of them after removes; \S\ref{sec:cost} turns these counts into waits. Operation by operation, the source writes two lines for an insert and one for a remove; the ordering flushes write none for an insert into a free slot, two for an insert that extends a chain, and one for a remove. The saving is the insert into a free slot, and our discipline keeps the flush after every remove.

\paragraph{Checks.}
The source's buckets have three slots and the model's two. With three slots, in \HiThreeRuns further exhausted runs, the outcomes are the same: no violation of \odl under both ordering flushes, \HiThreeFirstPairOdl violating histories under the first alone, \HiThreeNoneOdl under none in the instance that extends a chain. The \HiDepStopped explorations that stopped at their cap visit one corner of their state spaces; repeated with three other search orders (\HiOrderRuns runs of \HiCapM million states), they again show no violation of \odl under both ordering flushes, and the same orders find the violations of the weaker disciplines in all \HiOrderPowerRuns runs made to check that they can. The two instances in which the crashed host restarts, the only ones in which a lock is taken over, are among those that stopped. Guo et al.\ report three missing flushes in P-CLHT's constructor and in the allocation of its table~\cite{cxlmc}; those paths precede the state our instances start from.

\section{An allocator}
\label{sec:allocator}

CXL-SHM is an allocator for memory shared by clients on different machines~\cite{cxlshm2023}. Each object carries a reference count. A client that crashes is recovered by a monitor, which releases the references the client held: its root references, the references it had attached to a queue and not yet passed, and, through a redo entry, a reference it was attaching or releasing when it crashed. Clients pass references through queues with one sender and one receiver. The sender attaches the reference to the slot at the tail and then advances the tail, and ``ownership of the reference recorded in the queue is transferred atomically from A to B at the point where A advances the tail offset of the queue''. Each attach and release of a reference writes and flushes a redo entry before the count changes, then advances and flushes the client's era word, a counter by which recovery tells whether a redo entry still applies. On top of these the allocation path flushes the new root reference. The authors add this flush ``to persist the modification because our current implementation is still based on CXL 2.0'', measure that it ``can occupy 27\%$\sim$50\% of the total time'', and write that it ``may not be required in a CXL 3.0 based implementation that implements hardware distributed cache coherence (e.g., with an eADR-like assistance to flush cache data at the time of node failure)''. In their account of the order of allocation, the root reference is linked to a block before the block is taken from its page's free list, so that a client that crashes between the two steps does not leak the block. That flush is the one we study.

We wrote allocation, reference attachment and release, sending, receiving and recovery as steps from the public source, with the monitor's recovery taking one memory access at a time beside a surviving client, and with objects that hold a reference to another object. A queue's receiver is recorded when the queue is created, as the paper describes. Every discipline below keeps the source's flushes of redo entries, of era words and of a fresh page; the disciplines differ in the flush of the root reference and in the flushes we add. The crash condition is safety for the survivor: it never holds, or reaches through a reference inside an object it holds, an object whose count is zero or whose page or segment has been freed. There are \AlInstances instances (\AlRuns runs with a crash or none, \AlStatesM million states); in \AlLeaveInstances of them a reference or a segment leaves the client that crashes. Of the \AlRuns runs, \AlStopped stopped at a cap of \AlCapUnsafeM million states, none of them under the ordering flushes.

\paragraph{The ordering pairs.}
The store that passes ownership is the advance of the tail. By \S\ref{sec:ordering} it depends on every line that recovery or the receiver will read through the reference. Which lines those are we found by exploration, not by reading: starting from a flush, before the tail advances, of every line that the client writes and recovery reads, we removed one line at a time and kept those whose removal made an instance unsafe. They are the line of the object sent, the lines of the objects it reaches through references (we explored chains of one and two references), the metadata of their page, the head of the client's list of pages when the page is new, and the client's era. Two more pairs do not involve a queue. Within an attach or a release, the advance of the client's era depends on the reference just written; we added this flush after an instance in which the sender releases its own reference showed the need. And the release of a segment depends on the head of the list of pages. The discipline we call the ordering flushes consists of these flushes, with two lines in the flush before the tail's advance that removal did not show to be needed (the line of the client's other object and its pointer to its queue). It does not include the flush of the root reference. The lines were fixed on the ten instances in which no object holds a reference. The four instances with such an object and the two with a chain of references were explored afterwards, and what the ordering flushes and a discipline with one line left out would give in them was predicted before those runs.

\paragraph{Results.}
With every modified line written back at the crash, the assumption under which the authors expect the flush to be unnecessary, no execution is unsafe under any discipline (\AlKeptRuns runs, every state space exhausted). The rest is with lost caches.

With the flush of the root reference and none of the ordering flushes, executions are unsafe in all \AlPubUnsafeInstances instances in which something leaves the client. In the first one found, the client allocates the first block of a page, sends it and crashes. The device has the queue's slot and tail and the object's count, and the page's metadata from before the allocation, whose free pointer still names the block. Recovery takes the block for a free one, sets its count to zero and clears the page; the survivor then receives the reference and holds an object in a cleared page. With neither that flush nor the ordering flushes, executions are unsafe in the same instances, in \AlNoneLaterRuns of them found only when the run was repeated with a cap of \AlNoneLaterCapM million states in place of \AlCapUnsafeM million.

\textbf{With the ordering flushes and without the flush of the root reference, no execution is unsafe} in any of the \AlInstances instances, and every one of these state spaces was exhausted, the largest at \AlOrdLargestM million states. With the line of the object that the sent object refers to left out of the flush before the tail's advance, executions are unsafe again in \AlSentOnlyUnsafeInstances of the \AlSentOnlyInstances instances, among these \AlInstances, that have such an object, each explored to \AlCapUnsafeM million states (\AlSentOnlyUnsafe unsafe executions in the first of them): the survivor reaches, through the object it received, an object whose count on the device is zero.

In the chains we explored, the dependency reaches as far as the references do. In \AlChainInstances further instances the sent object refers to an object that refers to a third. With the lines of all three in the flush before the tail's advance, no execution is unsafe; these state spaces did not fit in one run, and after single runs had stopped at their caps we exhausted them in \AlChainRuns runs that divide the executions by the point at which the client crashes. With the third object's line left out, executions are unsafe in both instances: the reference to the third object is on the device and the count that covers it is not, and the survivor reaches, through two references, an object whose count is zero.

When nothing leaves the client, no discipline has an unsafe execution, the one with neither the root-reference flush nor the ordering flushes included. A recovery with a deliberate error is caught in every instance that passes a reference.

So the flush of the root reference is neither sufficient nor necessary for the survivor's safety when a host crash loses the cache, given the flushes of redo entries and eras that the source also issues: it is the ordering flush of another pair, the page's free pointer after the root reference that names the block, before anything has left the client. That pair keeps a crashed client from leaking a block, which the survivor's safety does not ask for. What a design of this kind needs for the survivor's safety is a flush when ownership passes, of the lines the new owner will depend on.

\paragraph{What the ordering flushes do not give.}
Under every discipline some executions end with an object that keeps a positive count and has no holder, when the crashed client's count reached the device and its reference did not. Such leaks are what the pair of the root-reference flush guards against, and the ordering flushes for safety leave more of them than the source's flush does: in the first instance \AlLeakOrdFirst of \AlTermOrdFirst end states of the model leak under the ordering flushes and \AlLeakSrcFirst of \AlTermSrcFirst under the source's flush, and the order is the same in each of the \AlLeakSrcExhausted instances in which both state spaces were exhausted and a leak occurs. These are counts of distinct end states, not probabilities.

The two do different jobs, and a design can keep both. With the root-reference flush and the ordering flushes together, no execution is unsafe in any of the \AlBothRuns instances, every state space exhausted, and fewer end states leak than under the ordering flushes alone in each of the \AlLeakInstances instances that have a leak (\AlLeakBothFirst of \AlTermBothFirst in the first instance).

\paragraph{Scope.}
The public code registers a queue's receiver at its first receive, not at the queue's creation. With that registration our model has unsafe executions under every discipline, stored energy included, through an interleaving of the registration with the monitor's release of unreceived references in which no flush plays a part; the results above are for the registration the paper describes, which we adopted after the code's registration failed the control with stored energy, and they say nothing about flushes under the other. The model also follows the paper when the code's condition for applying a redo entry differs from the paper's. Guo et al.\ ran two of the allocator's benchmarks under their checker together with its recovery check, which looks for unfreed memory of crashed processes, and reported an unimplemented free procedure and a division by zero~\cite{cxlmc}. That check does not look for a survivor that holds a freed object. And the model gives every object a line of its own. In the source an object's block is its data and a 24-byte header, rounded up to a multiple of 16 bytes, and blocks lie back to back: an object with up to 8 bytes of data shares its line with one neighbor, and a block whose size is not a multiple of 64 bytes shares a line with a neighbor or spans two. A flush of a line carries whatever else is on it, so the results are for objects that each occupy one line alone.

\section{What the split costs and saves}
\label{sec:cost}

The split is a statement about which flushes are required, and it does not always remove flushes. The three designs show three outcomes; we express the waits as products of our counts and published unit times (\S\ref{sec:limits}).

\begin{figure}[t]
\centering
\begin{tikzpicture}
\begin{axis}[paper, scale only axis, width=46mm, height=30mm, xmin=0, xmax=850, ymin=0.4, ymax=6.6, ytick={1,...,6}, yticklabels={{ordering, every line touched}, {ordering, none touched}, {published code}, {ordering, every line touched}, {ordering, none touched}, {published code}}, ytick style={draw=none}, xlabel={wait for the device per operation (ns)}]
\draw[black!60, line width=5pt] (axis cs:532.0,6) -- (axis cs:788.0,6);
\draw[wonggreen, line width=5pt] (axis cs:153.1,5) -- (axis cs:226.8,5);
\draw[wongorange, line width=5pt] (axis cs:342.6,4) -- (axis cs:507.4,4);
\draw[black!60, line width=5pt] (axis cs:399.0,3) -- (axis cs:591.0,3);
\draw[wonggreen, line width=5pt] (axis cs:181.1,2) -- (axis cs:268.3,2);
\draw[wongorange, line width=5pt] (axis cs:290.1,1) -- (axis cs:429.7,1);
\node[anchor=east, font=\scriptsize\bfseries] at (axis cs:840,5) {inserts};
\node[anchor=east, font=\scriptsize\bfseries] at (axis cs:840,2) {half removes};
\draw[black!25] (axis cs:0,3.5) -- (axis cs:850,3.5);
\end{axis}
\end{tikzpicture}
\caption{Estimated wait for the device per operation of the hash index, for a memory device attached directly to the host. A bar spans the published range of the time of one flush of a modified line, \EsWDirectLo to \EsWDirectHi\,ns~\cite{hitchhiker2025}, multiplied by the lines the operation writes, counted on P-CLHT's own code (resizes left out). \emph{Touched} refers to the lines that the ordering flushes leave modified: if another host is the next to access such a line it waits for the write-back, charged here as one more flush. Top: inserts into an empty table. Bottom: half inserts and half removes.}
\Description{Horizontal range bars in two groups of three. Inserts: the published code waits 532 to 789 nanoseconds per operation; the ordering flushes 153 to 227 when no other host touches the lines left modified and 343 to 508 when another host touches every one. Half inserts and half removes: 399 to 592, 181 to 269, and 290 to 430 nanoseconds.}
\label{fig:cost}
\end{figure}
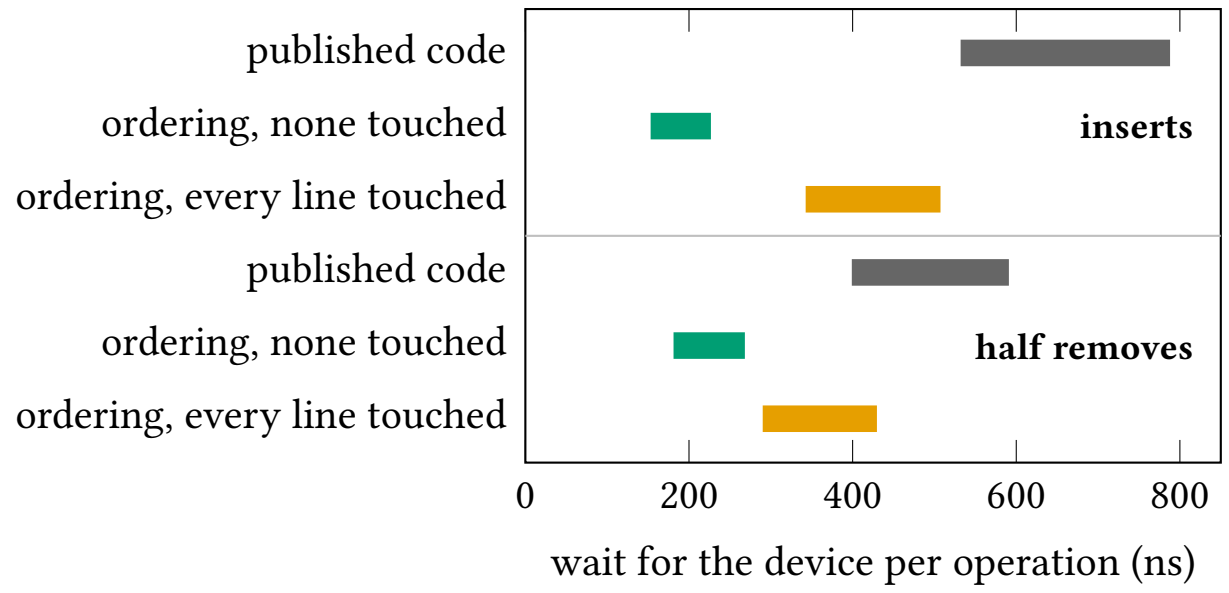

\paragraph{Unit times.}
An operation waits for two things. $W$ is a flush of one modified line to CXL memory. On six combinations of four processor generations with expansion devices attached directly to the host it takes \EsWDirectLo to \EsWDirectHi\,ns~\cite{hitchhiker2025}; an independent measurement gives \EsCxlZeroFlushDev\,ns, against \EsCxlZeroFlushHost\,ns for the host's own memory~\cite{cxl0}. On prototypes of pooled and of shared memory it takes \EsWProtoLo to \EsWProtoHi\,ns~\cite{hitchhiker2025}. For a device behind a switch, which adds at least \EsSwitchAdds\,ns to a round trip~\cite{octopus2026}, we derive \EsWSwitchLo to \EsWSwitchHi\,ns; no one has measured it. $D=\rho W$ is what an access waits, beyond an ordinary miss, when it forces another host's write-back. No one has measured that either. The premise puts a write-back inside it, so $\rho\ge 1$. A model of the protocol counts six remote message delays for a write when the owner holds the line modified and four when the owner holds the line clean~\cite{c3coherence2026}, and a simulator gives about \EsXerxesDirtyLo\,ns against \EsXerxesCleanLo to \EsXerxesCleanHi\,ns for a write that must snoop a peer device~\cite{xerxes2026}. We report $\rho=1$ and $\rho=2$.

\paragraph{Fewer.} The hash index flushes for durability on every update, and the ordering pairs ask for a flush only when a chain grows or a pair is removed: \HiLinesShareLoadPct and \HiLinesShareSteadyPct percent of the lines the source writes in the two phases. A remove writes one line under both disciplines and an insert that extends a chain writes two, so what the ordering pairs save is the two lines of an insert into a free slot. These are the writes an operation issues itself, by a flush or a non-temporal store. A line it leaves modified reaches the device later, at an eviction or when another host accesses it, and that host then waits for the write-back.

Charge one $W$ for each line an operation writes, including a non-temporal store, for which we found no measurement on CXL memory. An insert into a free slot then saves two waits and leaves one line modified. Let $p$ be the share of those lines that another host is the next to touch; each costs that host $D$. Operations wait less under the ordering flushes than under the published code exactly when $p\rho<\EsHiBreakEven$, whatever $W$ is, so at $\rho\le 2$ they never wait more. On a directly attached device an insert into an empty table waits \EsHiSrcNsLoLoad to \EsHiSrcNsHiLoad\,ns under the published code, and \EsHiOrdNsLoLoad to \EsHiOrdNsHiLoad\,ns under the ordering flushes when no other host touches the lines left modified, \EsHiSavePctLoad percent less; when another host touches every one of them and $\rho=1$, the saving is \EsHiSaveHalfPctLoad percent (Figure~\ref{fig:cost}). For half inserts and half removes the two figures are \EsHiSavePctSteady and \EsHiSaveHalfPctSteady percent. The measurements show two lines flushed together taking about as long as one~\cite{hitchhiker2025}; the charge counts every line, both for the two lines that an extending insert flushes in one step under the ordering pairs and for the two lines that an insert of the published code writes. The flush after a remove could be deferred to the next insert of the same key into another line; we did not implement that, and it would have the host remember which of its removals are unflushed.

\paragraph{A share of the stores.} The lock table follows a design that has no flush. Its first ordering flush follows each exclusive acquisition: \LkExclMin to \LkExclMax flush steps per committed transaction, which is \LkExclShareMinPct to \LkExclShareMaxPct percent of the stores to lock records (Figure~\ref{fig:locks}b). A flush is not the only wait: a line left modified is written back when another host next asks for it, and that host waits. With the unit times above for a directly attached device and $\rho$ from 1 to 2, the lock records cost \EsLkNoneUsLo to \EsLkNoneUsHi\,$\mu$s of waiting per committed transaction with no flush, across the \EsLkMixes mixes with eight records per line, and \EsLkExclUsLo to \EsLkExclUsHi\,$\mu$s under the first rule; the rule adds \EsLkExtraOneUsLo to \EsLkExtraOneUsHi\,$\mu$s at $\rho=1$ and \EsLkExtraTwoUsLo to \EsLkExtraTwoUsHi\,$\mu$s at $\rho=2$. Appendix~\ref{app:locks} compares the disciplines as a function of $\rho$.

\paragraph{Elsewhere, and not fewer.} The allocator's source flushes each new root reference, once per allocation. The ordering flushes we add come once per reference passed, in one step that names the lines the reference depends on, and once each time the client attaches or releases a reference, when the reference is flushed before the client's era advances. They are not fewer than the flush of the root reference, which a design that also wants its protection against leaks keeps (\S\ref{sec:allocator}), and we did not count them on a workload.

\paragraph{Which hardware.} Documents for two coming server processors list CXL 3.0 and CXL 3.1 among their interfaces~\cite{inteldmr2026,amdepyc9006wp2026}; neither says whether that includes coherence across hosts. Huang et al.\ write that ``There are currently no physical CXL devices that support inter-host cache coherence''~\cite{tigon2025}, and a talk on support for back-invalidation in Linux notes that ``Hardware is catching up with CXL3 IP''~\cite{bueso2025lpc}. On hardware that keeps caches coherent across hosts, $W$ is at least a round trip to the device and $D$ at least $W$; the estimates above move with the unit times and the break-even $p\rho<\EsHiBreakEven$ does not.

\section{Related work}
\label{sec:related}

\paragraph{Ordering before durability, on one machine.}
The ordering rules themselves are old. Soft updates orders the dependent metadata writes of a file system by three rules, among them ``Never point to a structure before it has been initialized'' and ``Never re-use a resource before nullifying all previous pointers to it''~\cite{mckusick1999}, and tracks the dependencies so that the order holds without synchronous writes~\cite{softupdates}; several of the pairs of \S\ref{sec:locktable} to \S\ref{sec:allocator} are instances of these two rules: the link and the bucket it names, the key stored again and its earlier removal, the reference passed and the object it names. The same paper separates the write that orders from the write that makes durable: after the write that orders an inode before its directory entry, ``most BSD fast filesystem implementations also immediately write the directory block before the system call creating the file returns'', a second write that it calls ``not strictly necessary'' and that ``ensures that the filename is on stable storage''~\cite{mckusick1999}. Memory persistency separates the order in which stores become visible from the order in which they persist~\cite{pelley2014}. Its persist barriers order persists, and its buffered strict persistency lets the persistent state lag behind execution, with a ``persist sync'' to make it catch up before a side effect that must not precede it. Optimistic crash consistency gives a file system two primitives, osync and dsync, ``that decouple ordering of writes from their durability''~\cite{chidambaram2013}. External synchrony owes durability only to an observer outside the machine: the system holds output back until the modifications it depends on are on disk~\cite{nightingale2006}. These are earlier statements of the split we use: ordering is one obligation, and synchronizing persistence with what an observer has seen is another. All of them concern one machine, whose whole cache is lost at a crash. In a pod the observer that matters is a surviving host, and under the premise the fabric performs the synchronization for every line that host touches (Lemma~\ref{lem:obs}), and for that line only; what remains for software is the ordering. Among correctness conditions the nearest to \odl is buffered durable linearizability~\cite{izraelevitz2016}, which also lets a crash remove completed operations; \S\ref{sec:model} says how the two differ.

\paragraph{Models and checkers for pods.}
Assa, Lumme, B\"urgi, Friedman and Lahav give the first programming model for concurrent programs over CXL, an operational semantics with independent machine crashes, and prove that an adaptation of FliT~\cite{flit} makes any linearizable object durably linearizable under partial crashes~\cite{cxl0}; that transformation flushes for durability, which Proposition~\ref{prop:lower} shows \dl cannot avoid. They note that with the write-back the transformation's counters can be kept in local memory, and they name as future work that ``relaxing durability semantics has generally been shown to be beneficial for performance \ldots{} and can be explored here as well''; this paper explores one such relaxation. Guo, Truong and Demsky build a model checker for crash-consistency bugs in programs over CXL shared memory and use the write-back that a load by one host forces on a line owned by another~\cite{cxlmc}. Their checker finds missing flushes in given code; it does not say which flushes a design could drop. Guo, Luo and Demsky build a tool that inserts flushes and fences automatically so that a program is robust, a criterion about the order in which stores persist, and state that the tool applies to CXL shared memory~\cite{guoluo2025}. Their analysis treats every cached line as lost at a crash, and in their example robustness requires a flush of a line after another thread has read it; when that thread runs on another host, the premise has already put the line on the device.

\paragraph{Pod systems and host failure.}
Published designs for pods answer a host crash in one of three ways. Tigon treats the failure of any component as a failure of the whole system~\cite{tigon2025}. Lupin tolerates partial failures with recoverable locks, assuming a global persistent flush~\cite{lupin}. CXL-SHM flushes on its allocation path and recovers by reference counts~\cite{cxlshm2023}, and XTRA flushes a redo log and its updates at commit and has survivors scan for a crashed host's locks~\cite{xtra2026}. None of them states which of its flushes a host crash requires; \S\ref{sec:allocator} measures one of them against the split. Other designs avoid the question by not relying on coherence between hosts: Borges gives every shared metadata word a single writer because ``current CXL pods lack inter-host cache coherence''~\cite{borges}, and Cxlalloc works with full, limited or no coherence~\cite{cxlalloc}.

\paragraph{Protection below the software.}
Other work removes the loss itself: by extending the coherence transaction of every write so that replicas log it in hardware~\cite{diktamo}, or by an object store for function-as-a-service workloads built so that the crash of a server that shares an object does not block the others~\cite{apta2023}. The recoverable distributed shared memory systems of the 1990s integrated checkpointing or logging with the coherence protocol~\cite{wufuchs1990,icare1995}. These change the fabric or the runtime. We ask what software must do on a fabric that offers the premise and nothing more.

\section{Limitations}
\label{sec:limits}

\paragraph{The premise.} Every result assumes a fabric that writes a modified line back before another host's access to it is served. That CXL's hardware-coherent shared regions are such a fabric is our reading of the specification (\S\ref{sec:model}), which states the order of the write-back and the response in the description of one flow; it is not a measurement. Assa et al.\ report that no processor or pool device implements these flows~\cite{cxl0}, and no hardware that keeps caches coherent across hosts was available to us. An emulator whose authors write that it implements the full CXL 3.0 coherency protocol has transfers between hosts that ``allow one host to service a cache miss from another host's cache without indirecting through main memory''~\cite{ocean2026}; on hardware that behaved so the premise would fail, and the lock table of \S\ref{sec:locktable} shows what is then lost.

\paragraph{The crash.} The model's crash erases one host's cache and leaves the device, and the other hosts' access to it, as they were, at once. On our reading the specification gives the first part for a shared region on a multi-logical device behind a switch, on which a reset of the crashed host's logical device releases that device's share of the region and no other's, and not on a multi-headed device, on which a conventional reset of one head releases the shared capacity from every host~\cite[\S9.13.3]{cxl30spec}. It does not give the second: while a snoop is pending the device blocks requests for the line~\cite[\S3.3.3]{cxl30spec}, so an access to a line that the crashed host held can wait for a response that will not come, as Patil et al.\ observed of CXL 3.0~\cite{apta2023}. We found no rule in revision 3.0 that ends the wait. In revision 4.0 the memory device of a port-based-routing fabric can notify the requesting host that a back-invalidate request received no response, as a failed memory request~\cite[Table 8-240]{cxl40spec}, which is not the access served from the device that the model's crash step gives. Our results concern what the device holds, not how soon it can be reached. A third behavior has been described, in which a line that a crashed host held modified is reported as poisoned and an access to it raises an exception: Assa et al.\ discuss it as a variant of their model, the checker of Guo et al.\ has it as an option, and Guo, Luo and Demsky report that it was ``not yet decided how CXL will expose such failures to the software layer''~\cite{cxl0,cxlmc,guoluo2025}. Under poisoning every line that the crashed host held modified at the crash is unreadable to the survivors, whatever the device holds, so the flushes that our results show a host crash does not require would buy availability there, and our results do not cover it.

\paragraph{The theorems.} Theorem~\ref{thm:kept} and Proposition~\ref{thm:needed} concern one ordering pair, and Theorem~\ref{thm:commit} concerns operations that take effect in one write and whose prepared lines no second operation is given; the update paths of the three designs write several times, and for them the crash condition is tested (\S\ref{sec:ordering}). For the allocator, reading gave the stores that pass ownership and exploration gave the lines they depend on.

\paragraph{The models.} Each study explores a model that we wrote of another group's design, on instances with two or three hosts and one or two crashes. For the hash index \HiDepStopped of the \HiInstances explorations under the ordering flushes stopped at \HiCapM million states, its model has two slots per bucket in all but \HiThreeRuns runs and no resizing, its allocator of buckets is our stand-in, and it has a lock that records its holder and a flush at each host's start that the source lacks (\S\ref{sec:hashindex}). A survivor that waits for the lock of a host that never restarts is not counted as a violation, and the instances in which a lock is taken over were not exhausted. For the allocator the registration of a receiver follows the paper and not the public code (\S\ref{sec:allocator}), an object holds at most one reference, and a crash of the monitor during recovery is not modeled. Its model gives every object and every root reference a line of its own. In the source, root references are 16 bytes, four to a line, and object blocks come in sizes 16 bytes apart and lie back to back, so a block whose size is not a multiple of 64 bytes shares a line with a neighbor or spans two lines; the model represents neither, and its results are for objects that each occupy one line alone. The lock-table runs execute lock requests, not transactions, and do not run a recovery.

\paragraph{What was not measured.} Flushes were counted, not timed: the times in \S\ref{sec:cost} are products of our counts and of unit times that others measured on devices without coherence across hosts. We know of no measurement, on hardware, of an access that waits for another host's write-back, and none of a flush through a switch.

\paragraph{What the condition gives up.} Under \odl a host that crashes may lose operations it had completed. A system whose clients outlive the host, or which acknowledges work outside the pod, needs the durability flush for whatever it acknowledges.

\section{Conclusion}
\label{sec:conclusion}

A flush in a design for persistent memory either orders two lines or makes an operation durable. On shared memory whose fabric writes a modified line back before another host may access it, a host crash requires the first kind and not the second, when what a crashed host alone has seen may be lost: we proved this split for operations that take effect in one store, and for a lock table, a hash index and an allocator, whose operations store several times, we found no violation under the ordering flushes alone in the instances explored on such a fabric.

Two implications are untested. Tools that insert or check flushes for persistent memory could take the fabric's write-back into account and place flushes by dependency. And a fabric that guarantees the write-back gives software more than coherence: it gives every surviving host a device that already holds what that host has seen.

\begin{acks}
This work was supported in part by the National Science Foundation under award 2446084.
\end{acks}

\bibliographystyle{ACM-Reference-Format}
\bibliography{references}

\appendix
\section{Details of the lock-table runs}
\label{app:locks}

The lock-table runs of \S\ref{sec:locktable} execute lock requests on the model, not transactions, and their units are the model's. Eight hosts run three workers each. A worker's transaction issues its lock requests one at a time under strict two-phase locking; a refused request aborts the transaction, which releases its locks and starts over, and a commit releases them. At each step one worker, drawn uniformly, takes its next action. This fixes how long a lock is held in steps of the other workers and is not a model of time. A tuple has a lock record, which holds its exclusive holder and its shared holders, from the first access by a host other than its owner; before that its locks are in the owner's private memory and outside the model. Every access to a record, a refused request included, is one read-modify-write of the record's line by the accessing host, since Tigon's record carries a latch that a thread takes before it accesses a tuple~\cite{tigon2025}. Each host's cache keeps the 4,096 record lines it accessed last and writes a line back upon evicting it, and no other write-back happens without a snoop. A further \LkReplicaRuns runs with caches of 256 and of 65,536 lines, made with one record per line and with eight in order of creation, give the same counts at every sampled crash as the runs with 4,096 in all \LkCachePairs pairs compared, and are not counted in the totals of \S\ref{sec:locktable}.

The mixes and their settings are taken from Tigon's evaluation, and each transaction is reduced to a simplified set of lock requests on existing rows: no transaction inserts a row, customers are chosen by identifier, a delivery locks one customer in each district, and a stock-level transaction locks its district and 20 stock rows. TPC-C has 24 warehouses, three per host, and its standard mix of five transactions; an order line is supplied by another warehouse with probability 0.01 or 0.0875, which gives 10 or 58 percent of the orders a remote line, and a payment's customer belongs to another warehouse with probability 0.15 or 0.90. These are the 10/15 and 58/90 mixes of Figure~\ref{fig:locks}: the first is TPC-C's default, and the second is near the upper end of the range Tigon evaluates. Two of the 23 other warehouses are on the same host, and an access to one of them involves no lock record. The key-value mix is YCSB as Tigon configures it, with 2.4 million tuples partitioned by range over the hosts and ten operations per transaction on Zipfian keys, with skew 0.7 or 0.99, with 95 or 50 percent of the operations reads, and with 10 or 100 percent of the transactions placing five of their operations in another host's partition. The most popular key of a partition draws \LkColdKeyPct percent of its accesses at skew 0.7 and \LkHotKeyPct percent at skew 0.99. A run commits 200,000 transactions. After the first fifth of it, every 2,500 steps one host is drawn and the lock table that its crash would leave is computed without changing the run: every line that the host holds modified takes the device's value, and every other line keeps its own. Against the locks in force we count the locks of running transactions on other hosts that this table lacks, the locks it shows that transactions on other hosts had released, and the crashed host's own locks that it shows and that it lacks.

A lock record is 8 bytes, so a line holds one record or eight, and with eight the result depends on which records share a line. In the layout \emph{in order of creation} each owner fills its lines with records as they are created, so that a line holds records of one owner whose first uses by a second host were consecutive. In the layout \emph{at random} every record that the first layout creates in the same run exists from the start, and each owner's records are placed in a uniformly random order. A third layout has every record present from the start in the slot that the first layout gives it. In the second and third layouts the owner's own accesses to those tuples go through their records from the first step. We added these two layouts after seeing the results of the first, expecting the creation of records during a run to account for its shares. It does not: the third layout, which creates none, differs from the first by at most \LkPlacementGapMaxPts percentage points in any mix, and against the random layout, from which it differs in placement alone, its share is higher in every mix, by a paired difference over the seeds whose 95 percent interval lies above \LkContrastMinLoPts points. With one record per line a survivor's lock is lost only when the crashed host was the last to access the very record on which another host's transaction holds a lock, and the mean share is at most \LkTpccOneUp percent for TPC-C (\LkTpccHiOneHits of \LkTpccHiOneCrashes sampled crashes at 58/90 and none of \LkTpccLoOneCrashes at 10/15), at most \LkColdOneUp percent at skew 0.7, and \LkHotOneDn to \LkHotOneUp percent at skew 0.99. Eight records per line placed at random add at most \LkRandGapMaxPts points to that. Eight per line in order of creation give \LkTpccHiUse percent for TPC-C at 58/90 and \LkAllUseDn to \LkAllUseUp percent for the key-value mixes whose transactions all cross hosts. A table in order of creation starts empty, and a record receives on average between \LkStoresPerRecordMin and \LkStoresPerRecordMax stores in a run, by mix. Tigon also moves tuples back out of the shared memory under a memory budget, which the runs do not model.

Without forwarding no sampled crash loses a lock of a survivor or brings back one that a survivor released. Of the locks that the crashed host's own transactions hold in lock records, at least \LkAbsentMinPct percent are lost in every mix and layout. What that loss can do, and which flushes prevent it, we explored exhaustively on one lock and the two data lines it guards, all three in the model's memory. A writer takes the lock exclusively, stores to both lines and releases; a reader takes the lock, shared or exclusive, loads both lines and releases; a refused request ends its transaction. The explorer runs every interleaving of the threads' steps with a crash of any host at any point, within a budget of one or two crashes, and leaves each line that the crashed host held modified with any value that a silent write-back could have put on the device (\S\ref{sec:method}). A crash is offered while some thread has a step left, and a writer flushes its two data lines in one step. The property is that a thread that is granted the lock and loads both lines reads the two values of one transaction, or the two initial values, and that a thread that releases finds its own lock in the lock word. The \LkExInstances instances are nine programs, the first of them explored with one crash and again with two; they have two or three hosts and two or three threads: a writer against an exclusive or a shared reader, against two shared readers, and against a reader that reads twice; two writers on two hosts, in two threads of one host, and one after the other in one thread; writers that load both lines before they store to them; and a writer and a reader on one host. Every state space was exhausted, at between \LkExStatesMin and \LkExStatesMax states. With no crash, no instance violates the property under any of the disciplines below, and without forwarding no thread ever fails to find its own lock.

With no flush, every instance has an execution that violates the property: one data line reaches the device and the lock does not, the host crashes, and a survivor is granted the lock and reads one new value and one old. With a flush of the lock's line after the exclusive acquisition and no other, every instance has one again: the release and one data line reach the device, and the other data line does not. With a flush of the two data lines before the release and no other, every instance has one, as with no flush. With both flushes no execution of any instance violates the property, and neither does any of \LkExStressRuns random schedules on four hosts with two threads each, two transactions per thread and up to three crashes; the three weaker disciplines violate it in \LkExStressNone, \LkExStressLock and \LkExStressData of as many. In these programs no reader flushes, and no shared acquisition, refused request or release is followed by a flush of the lock's line. With both flushes and pass-dirty forwarding, every instance has a violating execution: a lock that its holder never flushed, such as a shared reader's, is lost with a host that had only requested it, and in \LkExFwdTornInstances of the instances a thread also reads a pair that no one transaction wrote. Under forwarding the explorer offers only some of the values that a write-back could leave, which can hide violations and not create them. The explored programs hold one lock, upgrade none and create no lock record; in the runs above a record's creation is followed in the same step by the access of the host that asked for it.

The first of the two flushes is the one that Figure~\ref{fig:locks}b counts. Exclusive acquisitions, upgrades included, are \LkExclShareMinPct to \LkExclShareMaxPct percent of the stores to lock records, by mix, and the other stores are shared acquisitions, refused requests, releases and the creation of records. The second is the flush by which a transaction puts its updates on the device before it releases. A design whose data lies outside the coherent region issues it in any case, since there a store reaches the other hosts only through the device; for data inside the region it is one more flush step for each transaction that writes. The lock-table runs do not model the guarded data and do not count this flush. The second rule applies to every release, so a transaction that aborts after a store has to have the lines it stored to on the device before it releases; the explored programs do not abort after a store.

Flush steps are not all that a discipline waits for. A host that leaves a lock's line modified has it written back when another host next asks for the line, and that host waits: a dirty miss. We charge a flush step that writes a line back one wait $W$, and a dirty miss $D = \rho W$ with $\rho \ge 1$, since the miss contains the write-back. We repeated the \LkFlRuns runs that create records at their first use, with one record per line and with eight, under four disciplines that see the same lock requests: no flush; a flush after each exclusive acquisition, which is the first rule; one flush step at the commit that names the lines of the records the transaction holds exclusively, which is the first rule for a design that defers its stores to the commit; and a flush after every store to a lock record. Per committed transaction, no flush leaves \LkForcedNoneMin to \LkForcedNoneMax dirty misses on lock records, by mix and layout. A flush after every store leaves none and issues \LkStoresMin to \LkStoresMax flush steps, each of which writes a line back, so it costs fewer waits than no flush only when $\rho$ exceeds the number of stores per dirty miss, which is between \LkFlBeAllMin and \LkFlBeAllMax. A flush after an exclusive acquisition removes a dirty miss only when another host is the next to access the line: it removes fewer than \LkFlSavedUp dirty misses per flush, and fewer than \LkFlSavedOneUp with one record per line. The first rule therefore costs more waits than no flush at every $\rho$ below \LkFlBeExclDn, and it leaves \LkFlExclKeptMinPct percent or more of the dirty misses of no flush (\LkForcedExclMin to \LkForcedExclMax per committed transaction). The flush step at the commit removes no dirty miss, because the releases that follow it store to every line it named; there are \LkFlBatchStepsMin to \LkFlBatchStepsMax such steps per committed transaction, each naming at most \LkFlBatchMaxNamed lines. At $\rho = 1$ no flush waits least and a flush after every store most in every mix and layout; at $\rho = 2$ a flush after every store waits least in \LkFlAllLeastAtTwo of the \LkFlPairs combinations of mix and layout. With a flush after each exclusive acquisition, no exclusive lock of the crashed host is missing from the lock table that would remain at any of the \LkFlCrashes sampled crashes, and with a flush after every store no lock of the crashed host is.

Under both rules the lock of a host that crashed between its exclusive acquisition and its release stays in the lock word, and the property holds in the explored programs because no thread is granted that lock again. Releasing it is the work of a recovery, which these programs do not have. A recovery that released the lock without first rolling the crashed holder's update forward or back would expose the two lines as the crash left them; rolling forward needs a log that is on the device before the update's first store, as XTRA's redo log is~\cite{xtra2026}.

These runs drive the model with lock-request patterns rather than a database with recovery, and they model the lock fields of a record alone; Tigon's record also carries the bits of its software cache coherence, whose stores are not modeled. XTRA's protocol is optimistic and holds locks only while a transaction commits~\cite{xtra2026}; it was not run. Forwarding stands for coherence protocols whose tables let a snooped cache supply a modified line to the requester.

\end{document}